\documentclass[prd,%
preprint,
tightenlines,superscriptaddress,showpacs,
nofootinbib,eqsecnum,amsfonts,amsmath,amssymb]{revtex4} 

\usepackage{bm} 
\usepackage{graphicx}

\allowdisplaybreaks
\begin{document}

\title{Dipolar Particles in General Relativity}

\author{Luc Blanchet} \email{blanchet@iap.fr}
\affiliation{${\mathcal{G}}{\mathbb{R}} \varepsilon{\mathbb{C}}{\mathcal{O}}$,
Institut d'Astrophysique de Paris --- C.N.R.S., Universit\'e Pierre \& Marie
Curie, 98$^{\text{bis}}$ boulevard Arago, 75014 Paris, France}

\begin{abstract}
The dynamics of ``dipolar particles'', \textit{i.e.} particles endowed
with a four-vector mass dipole moment, is investigated using an action
principle in general relativity. The action is a specific functional
of the particle's world line, and of the dipole moment vector,
considered as a dynamical variable. The first part of the action is
inspired by that of a particle with spin moving on an arbitrary
gravitational background. The second part is intended to describe, at
some effective level, the internal non-gravitational force linking
together the ``microscopic'' constituents of the dipole. We find that
some solutions of the equations of motion and evolution of the dipolar
particles correspond to an equilibrium state for the dipole moment in
a gravitational field. Under some hypothesis we show that a fluid of
dipolar particles, supposed to constitute the dark matter, reproduces
the modified Newtonian dynamics (MOND) in the non relativistic
limit. We recover the main characteristics of a recently proposed
quasi-Newtonian model of ``gravitational polarization''.
\end{abstract}

\date{\today}

\maketitle

\section{Introduction}\label{secI}
\subsection{Astrophysical motivation}\label{secIA}
It is recognized (see Refs.~\cite{Milgrev, SandMcG02, Bekrev, Sandrev}
for reviews) that the modified Newtonian dynamics or MOND, which has
been proposed by Milgrom~\cite{Milg1, Milg2, Milg3} as an alternative
to dark matter, works extremely well at predicting the form of
rotation curves of galaxies from the observed distribution of stars
and gas. In addition MOND explains naturally the relation between the
luminosity of galaxies and their asymptotic rotation velocity --- the
so-called Tully-Fisher law~\cite{TF}. Though there might be some
examples where MOND does not fully account for the observed kinematics
of galaxies~\cite{Gentile}, overall the fit achieved by MOND of the
rotation curves of most galaxies is very impressive and calls for a
possible physical explanation. On the other hand we know that the mass
discrepancy of clusters of galaxies is not completely accounted for by
MOND~\cite{GerbalDurret}, and that at the cluster scale there is still
empirical evidence for unseen dark matter~\cite{Clowe}.

In the usual interpretation, MOND is viewed as a modification of the
fundamental law of gravity, without the need of dark matter. Several
relativistic extensions of MOND, sharing this view of modifying the
sector of gravity, postulate the existence of supplementary fields
associated with the gravitational force, in addition to the metric
tensor of general relativity~\cite{BekM84, BekS94, Sand97, Bek04,
Sand05}. So far the most developped of the relativistic MOND theories
is the tensor-scalar-vector theory of Bekenstein \&
Sanders~\cite{Sand97, Bek04, Sand05}.

At this stage we seem to face two alternatives to the issue of dark
matter:
\begin{itemize}
\item[(1)] Either accept the existence of cold dark matter particles,
\textit{e.g.} predicted by super-symmetric extensions of the standard
model of particle physics (see~\cite{BHS05} for a review). However
these particles are yet to be discovered, and the simplest models of
cold dark matter fail to reproduce in a natural way the flat rotation
curves of galaxies~\cite{SandMcG02}.
\item[(2)] Or postulate an alteration of our fundamental theory of
gravity (namely MOND and its relativistic extensions). But the
motivation for altering the law of gravity is \textit{ad hoc}, and one
can argue that the relativistic MOND theory~\cite{Sand97, Bek04,
Sand05} does not in fact explain the phenomenology of MOND --- at
least untill the extra fields find a well-motivated explanation coming
from fundamental physics.
\end{itemize}
\noindent
In the present paper, following~\cite{B06mond}\,\footnote{Hereafter
Ref.~\cite{B06mond} will be referred to as paper~I.}, we propose a
third alternative. Namely we:
\begin{itemize}
\item[(3)] Keep the standard law of gravity, \textit{i.e.} general
relativity and its Newtonian limit, but we add to the distribution of
ordinary matter some specific form of dark matter in such a way as to
naturally explain MOND. The dark matter consists of ``polarization
masses'' associated with a medium of dipole moments aligned in the
gravitational field of ordinary matter. (But, in this paper, we shall
leave aside the problem of clusters of galaxies~\cite{GerbalDurret,
Clowe}.)
\end{itemize}

Our basic motivation is that MOND can be naturally interpreted as
resulting from an effect of gravitational polarization of a dipolar
medium. Paper~I argued, on the basis of a simple quasi Newtonian
model, that the polarization tends to enhance the magnitude of the
gravitational field of ordinary galaxies in a way consistent with
MOND. This effect constitutes the gravitational analogue of the
electric polarization of a dielectric material (whose atoms can be
modelled by electric dipoles) by an applied electric
field~\cite{Jackson}. Thus the phenomenology of MOND results from the
non-standard influence of the dark matter on ordinary matter.

Arguably the proposal (3) does not really provide an explanation for
MOND. Indeed the models of paper~I and this paper are only
\textit{effective} --- the fundamental nature of the dipolar particles
(\textit{i.e.} their ``internal'' structure) will not be
elucidated. However the proposal is conceptually simple and fits
naturally with the MOND phenomenology of the flat rotation curves and
the Tully-Fisher empirical relation. In addition the model cannot be
said to be \textit{ad hoc} because it invokes the (gravitational
analogue of the) well-known physical mechanism of polarization by an
external field.

The quasi-Newtonian model of paper~I is based on a microscopic
description of the dipole moment using negative gravitational-type
masses (or gravitational charges). As a result the motion of dipolar
particles in this model violates the equivalence principle. In the
present paper we elaborate a model of dipolar particles and
gravitational polarization in the standard general relativity
theory,\,\footnote{The concept of gravitational polarization at the
quadrupolar order in relativistic gravity theories has been
investigated in Ref.~\cite{Bel}.} without negative (passive)
gravitational masses and consistent with the equivalence
principle. Consequently, we shall find that the equations of motion of
the dipolar particles in the non-relativistic (NR) limit of the
present model are different from those of paper~I. However we shall
recover the main characteristics of the quasi-Newtonian model of
paper~I, and notably the interesting connection with the phenomenology
of MOND.

\subsection{Concept of dipole moment in general relativity}\label{secIB}
In theories satisfying the equivalence principle the mass dipole
moment of a mass distribution, say $\pi^i_\mathrm{g}=\sum
m_\mathrm{g}\,x^i$ (adopting a Newtonian picture), is proportional, by
equivalence between the gravitational and inertial masses,
$m_\mathrm{g}=m_\mathrm{i}$, to the position of the center of mass of
the system, $C^i=\sum m_\mathrm{i}\,x^i/M_\mathrm{i}$ (where
$M_\mathrm{i}=\sum m_\mathrm{i}$). Similarly, the current dipole
moment, $\mu^i_\mathrm{g}=\sum m_\mathrm{g}\,\varepsilon^{ijk}x^j
v^k$, is equivalent to the spin angular momentum $S^i=\sum
m_\mathrm{i}\,\varepsilon^{ijk}x^j v^k$. Defining the mass dipole
moment of a particle (supposed to be composed of some sub-particles)
is \textit{a priori} delicate because one faces the problem that
$\pi^i_\mathrm{g}=0$ in the center-of-mass frame where $C^i=0$ by
definition; however the current dipole moment $\mu^i_\mathrm{g}$
equivalent to the spin $S^i$ is admissible. Thus, while the notion of
a particle carrying a mass dipole moment seems to be possible only at
the price of violating the equivalence principle (like in the model of
paper~I), the notion of a particle endowed with a current dipole
moment or spin is perfectly legitimate.

Particles with spins have been the subject of many fundamental works
in general relativity~\cite{papa51spin, Traut58, Dixon, BOC75, BOC79,
BI80}, with applications to the problem of dynamics of spinning black
holes in binary systems~\cite{KWW93, K95, TOO01, FBB06spin,
BBF06spin}.  The general formalism has been encapsulated in an action
principle for a particle with spin moving on an arbitrary
gravitational background. The action, due to Bailey \&
Israel~\cite{BI80}, is derived by expressing the standard action for a
non-spinning particle (\textit{i.e.} the integral of the proper time),
as defined with respect to some ``eccentric'' world-line located at
the edge of some composite particle, in terms of the quantities
belonging to a different world-line, which is viewed as the
``reference'' or ``central'' world-line located at the center of the
particle. Crucial to the formalism is the vector separation between
the two world-lines, defined as the gradient of the geodesic distance
separating them.\,\footnote{The geodesic distance is nothing but the
world function in Synge's formalism~\cite{Synge}; it is a bi-scalar in
the general theory of bi-tensors~\cite{dWB60}.} The spin of the
particle is then given by the anti-symmetric product between the
latter separation vector and the particle's linear momentum. Hence the
separation vector is the ``lever arm'' associated with the spin
angular momentum, as defined with respect to the fiducial central
world-line. The point for our purpose is that such lever arm appears
essentially to be the analogue of a mass-type dipole moment. We thus
see the possibility of describing dipolar particles in general
relativity by starting from (some variant of) the Bailey--Israel
action~\cite{BI80}.

The previous paradox concerning the dipole moment which should vanish
in the frame of the center-of-mass will be solved because (roughly
speaking) there are two notions of dipole moments. The first moment is
the one we mentionned earlier, $\pi^i_\mathrm{g}$, but let us call it
now the ``passive'' dipole moment $\pi^i_\mathrm{p}$ to emphasize the
fact that the gravitational mass it contains is really the
\textit{passive} gravitational mass $m_\mathrm{p}=m_\mathrm{g}$
(\textit{i.e.} that mass which enters in the right-hand-side of the
law of motion, in factor of the gravitational field). Assuming as
before that the dipolar particle is composed of sub-particles, its
passive dipole moment $\pi^i_\mathrm{p}\sim\sum m_\mathrm{p}\,x^i$
will indeed be zero in the particle's center of mass, by equivalence
between the inertial and passive gravitational masses of the
sub-particles, $m_\mathrm{p}=m_\mathrm{i}$. However the dipole moment
we shall consider in this paper is different: this is the moment
parametrizing the dipolar part of the stress-energy tensor
$T^{\mu\nu}$ of the particles (notably its $00$ component or energy
density). In general relativity the stress-energy tensor is the source
for the gravitational field, so we can rightly refer to this moment as
the \textit{active} dipole moment. In conclusion, we shall have
$\pi^i_\mathrm{p}=0$ by the equivalence principle, but we still have
at our disposal the ``active gravitational'' version of the moment,
call it $\pi^i_\mathrm{a}$, which enters the stress-energy tensor of
the particles, say through a term of the form $\sim
-\partial_i(\pi^i_\mathrm{a}\delta)$ in the energy density at the NR
approximation (where $\delta$ is the Dirac function). In the following
we shall investigate a specific model for the relativistic dynamics
and evolution of particles endowed with a four-vector dipolar moment
$\pi^\mu\equiv\pi^\mu_\mathrm{a}$ of the active
type.\,\footnote{Hereafter we no longer mention that this moment can
be regarded as the active one. Note that the names we are giving here
and below for different types of masses and dipoles are useful for the
discussion but are not required in the present formalism, which is
entirely based on the action~\eqref{S}.}

By analogy with the model of paper~I we shall denote by $2m$ the total
inertial mass of the dipolar particle, equivalent to its total passive
gravitational mass, \textit{i.e.}
$M_\mathrm{i}=M_\mathrm{p}=2m$. However, this equivalence does not
mean that the motion of the dipolar particle is geodesic. Indeed,
motivated by paper~I, we shall introduce a force $F^\mu$, which will
be considered as ``internal'' to the dipolar particle, and is aimed at
``stabilizing'' the dipole moment imbedded in an exterior
gravitational field. The force $F^\mu$ has a non gravitational origin,
and will derive from a scalar potential function $V$ in the
action. Because of the presence of this force reflecting its internal
structure, the dipolar particle is not a test particle: its motion is
not geodesic, its four-acceleration is non zero; the particle can be
thought of as a ``rocket'', self-accelerated by the internal force
$F^\mu$. We shall find an approximate solution in which $F^\mu$
accelerates the particle in such a way as to compensate for the local
gravitational field, so that the dipolar particle stays essentially at
rest in a gravitational field (like in the model of paper~I).

The monopolar part of the stress-energy tensor defines what can be
regarded as the particle's active gravitational mass
$M_\mathrm{a}$. Intuition from paper~I would lead to expect that for a
dipolar particle $M_\mathrm{a}$ is zero, \textit{i.e.} the particle
does not generate any monopolar gravitational field. In the present
model we shall find that this is not possible: $M_\mathrm{a}$ cannot
be zero, and in fact we shall have $M_\mathrm{a}=V/c^2$, where $V$ is
the potential function from which derives the internal force
$F^\mu$. However, thanks to the explicit factor $1/c^2$ it contains,
$M_\mathrm{a}$ turns out to be very small since it vanishes in the NR
limit, \textit{i.e.}  $M_\mathrm{a}=\mathcal{O}(c^{-2})$ when
$c\rightarrow\infty$. Thus, the dipolar particle will indeed be
``purely dipolar'' in the NR limit.\,\footnote{Notice that
the fact that the particle carries some different active and passive
gravitational masses, $M_\mathrm{a}$ and $M_\mathrm{p}$, does not
contradict the principle of conservation of the linear momentum (and
the law of action and reaction). Indeed the particle's stress-energy
tensor contains in addition to the monopolar part a dipolar
contribution (parametrized by $\pi^\mu$) which ensures the
satisfaction of the usual conservation laws.}

Applying the standard general relativistic coupling to gravity, we
prove that the Einstein field equations reduce to a Poisson equation
with a dipolar source term in addition to the density of ordinary
matter, having the structure $\sim -\partial_i(\pi^i_\perp\delta)$ in
the limit $c\rightarrow\infty$ (where $\pi^i_\perp$ refers to an
appropriate orthogonal projection). We are then close to the MOND
equation; to recover MOND it suffices that the dipole moment be
aligned with the gravitational field, and polarized in a certain
way. In the present paper we shall find, under some hypothesis, some
solutions which correspond to gravitational polarization,
\textit{i.e.} we shall show that the polarization scenario yielding
MOND is consistent with our equations. There will be essentially a
one-to-one correspondence between the internal potential scalar
function $V$ entering the action and the Milgrom~\cite{Milg1, Milg2,
Milg3} function $\mu$ (which is linked in the interpretation of
paper~I with the gravitational susceptibility $\chi$ of the dipolar
medium by $\mu=1+\chi$).

The paper is organized as follows. We first deal with the relativistic
model based on the action presented in Sect.~\ref{secIIA}. The general
equations of motion and evolution, and the stress-energy tensor, are
derived in Sect.~\ref{secIIB}. In Sect.~\ref{secIIC} we restrict
ourselves to a particular solution which corresponds to some
``equilibrium'' state for the dipole moment. For that solution we
obtain in Sect.~\ref{secIIIA} the dynamics of dipolar particles in the
NR approximation where $c\rightarrow\infty$. By neglecting the tidal
gravitational field we find in Sect.~\ref{secIIIB} that when the fluid
of dipole moments is at rest with respect to the ordinary matter
distribution the medium is polarized and the MOND equation follows.
Finally, we investigate in Sect.~\ref{secIV} (still in the NR limit)
the case of a stationary fluid of dipole moments aligned in the
central gravitational field of a point mass. Section~\ref{secV}
summarizes and concludes the paper. In the Appendix~\ref{appA} we show
on general grounds how to vary a dipolar action functional.

\section{Relativistic dynamics of dipolar particles}\label{secII}
\subsection{Action principle}\label{secIIA}
The starting point of our approach is the formalism for spinning
particles in general relativity~\cite{papa51spin, Traut58, Dixon,
BOC75, BOC79, BI80} and especially the Bailey--Israel
action~\cite{BI80}. In this formalism, the equations of motion of the
spinning particle (moving on an arbitrary gravitational background)
are derived from the extremum of the action with respect to an
infinitesimal displacement of the particle's world-line; however the
spin variable itself is not varied.  Instead, one postulates some
condition for the evolution of the spin, called the spin supplementary
condition (SSC), which essentially tells that the spin is parallel
transported along the particle's world-line (see
\textit{e.g.}~\cite{FBB06spin, BBF06spin}). In the present paper we
shall interpret the ``lever arm'' associated with the spin angular
momentum, in the formalism of spinning particles, as the mass-type
dipole moment carried by the particle. However we shall not impose
some analogue of spin supplementary condition. Instead we shall
promote the dipole moment as a \textit{dynamical} variable, which will
have to be varied independently of the variation of the particle's
position.

Let the particle follow a time-like world line $x^\mu$ in a four
dimensional manifold, and being endowed with the dipole moment four
vector $\pi^\mu$ (having the dimension of a mass times a
length).\,\footnote{Greek indices take the space-time values
$\mu,\nu=0,1,2,3$; Latin ones range on spatial values $i,j=1,2,3$; the
convention for the Riemann curvature tensor is
$R^\mu_{\,\,\,\nu\rho\sigma} = \partial_{\rho}\Gamma^{\mu}_{\nu\sigma}
- \partial_{\sigma}\Gamma^{\mu}_{\nu\rho} + \cdots$; the Lorentzian
signature is $-+++$; all factors in the speed of light $c$ and
Newton's constant $G$ are indicated throughout.} We propose that the
particle's dynamics in a prescribed gravitational field $g_{\mu\nu}$
be derived from the following matter action, consisting of three
terms,
\begin{equation}\label{S}
S = \sum\int_{-\infty}^{+\infty} d\tau
  \left[-c\sqrt{g_{\mu\nu}\left(2m\,u^\mu -
  \dot{\pi}^\mu\right)\left(2m\,u^\nu - \dot{\pi}^\nu\right)} +
  \frac{\dot{\pi}_\mu\,\dot{\pi}^\mu}{4m} -
  V\!\left(\frac{\pi_\perp}{m}\right)\right] \,.
\end{equation}
The sum goes over all the particles of this type. We denote the
particle's proper time by $d\tau=[-g_{\mu\nu}d x^\mu d
x^\nu/c^2]^{1/2}$, and its four-velocity by $u^\mu = d x^\mu/d\tau$
(which is normalized to $g_{\mu\nu}u^\mu u^\nu=-c^2$). The dynamical
variables are the space-time position $x^\mu(\tau)$ and the dipole
moment $\pi^\mu(\tau)$, both of them are function of the proper time.

The first term in Eq.~\eqref{S} is inspired by the known action for
the dynamics of spinning particles~\cite{BI80}, and describes the
inertial properties of the dipolar particle, \textit{i.e.} its
equations of motion in the gravitational field. The second term is a
kinetic-type term for the dynamical evolution of the dipole moment
itself, and serves to tell how this evolution will differ from
parallel transport. The third term of~\eqref{S} is made of a scalar
potential function $V$, supposed to describe, at some effective
``macroscopic'' level, a non-gravitational force that is internal to
the dipolar particle.

We denote by a dot the covariant derivative of the dipole moment with
respect to the proper time, namely
\begin{equation}\label{pidot}
\dot{\pi}^\mu \equiv \frac{D \pi^\mu}{d \tau} = \frac{d \pi^\mu}{d
\tau} + \Gamma^\mu_{\rho\sigma} u^\rho \pi^\sigma\,.
\end{equation}
More generally the dot will always refer to the covariant time
derivative $D/d\tau$. The potential function $V$ in the third term of
the action depends on the norm $\pi_\perp$ of the orthogonal
projection of the dipole moment perpendicular to the four velocity,
defined by
\begin{equation}\label{piperp}
\pi_\perp^\mu =\perp^\mu_\nu\pi^\nu\,,
\end{equation}
where $\perp^\mu_\nu\equiv\delta^\mu_\nu+u^\mu u_\nu/c^2$ is the
corresponding projection operator. The vector~\eqref{piperp} is
space-like and its norm is given by
\begin{equation}\label{normpiperp}
\pi_\perp = \bigl[g_{\mu\nu}\pi_\perp^\mu\pi_\perp^\nu\bigr]^{1/2} =
\sqrt{\perp_{\mu\nu}\pi^\mu\pi^\nu}\,.
\end{equation}
The fact that $V$ depends on the orthogonal projection $\pi_\perp^\mu$
rather than on $\pi^\mu$ itself is crucial for the present formalism.

The mass $m$ parametrizing the action~\eqref{S} is a certain mass
parameter associated with the dipole moment, and is such that
$x\equiv\pi_\perp/m$ represents the typical size of the dipolar
particle. The potential function $V$ will typically be [such as in
Eq.~\eqref{Vpiperp} below] a quadratic function of the particle's size
$x=\pi_\perp/m$. The mass $m$ can be viewed as the relativistic
analogue of the mass in the model of paper~I (we have already
discussed in the introduction some heuristic interpretation of $2m$).

An important feature of the action~\eqref{S} is that its first term is
given by the norm of a \textit{space-like} vector. It will be useful
to keep in mind this vector, so we introduce a special notation for
it:
\begin{subequations}\label{pmu}\begin{align}
p^\mu &= \frac{2m\,u^\mu -
\dot{\pi}^\mu}{\Lambda}\label{exprpmu}\,,\\\text{where}~~\Lambda &\equiv
\sqrt{-1 - \frac{u_\nu\dot{\pi}^\nu}{m c^2} +
\frac{\dot{\pi}_\nu\dot{\pi}^\nu}{4 m^2 c^2}}\label{Lambda}\,.
\end{align}\end{subequations}
This space-like vector, satisfying $p^2=+4m^2 c^2$, will \textit{not}
represent the linear four-momentum of the dipolar particle (otherwise
the particle would be a tachyon).\,\footnote{This is in contrast with
the Bailey--Israel action for a spinning particle~\cite{BI80}, which is
given by the norm of the particle's linear momentum --- a time-like
vector as usual.} As we shall see, since there are other terms in the
action, the linear momentum of the particle will differ from $p^\mu$,
and be normally time-like.

Because of the space-like character of $p^\mu$ we note that the
action~\eqref{S} makes no sense in the limit where the dipole moment
vanishes, since this limit corresponds to a particle with an imaginary
mass or tachyon. In this sense the dipolar particle described
by~\eqref{S} exists only through the existence of the dipole moment,
contrary to a particle carrying a spin, which is described both by a
spin and a mass, and reduces to an ordinary point mass in the absence
of the spin. [We are speaking of classical particles, whose spin is
given by the classical notion of angular momentum.] Here the mass
parameter $m$ is not independent from the dipole moment; it is such
that $\pi_\perp = m\,x$ where $x$ is the particle's size.

Unlike the case of a spinning particle~\cite{BI80}, no explicit
coupling to the Riemann curvature tensor has been introduced at the
level of the action~\eqref{S}. Recall that the Riemann tensor in the
Bailey--Israel action~\cite{BI80} comes from the particular way this
action is constructed, by conveying the point particle action from
some eccentric world-line of some composite particle to the physical
central world-line. As a result, if one varies this action with
respect to the spin, one obtains the same equation as with the
variation with respect to the position --- both equations containing
the same coupling to the Riemann curvature (at least at linear order
in the spin). In the present approach, by contrast, the dipole moment
will be considered as an independent dynamical variable, and we shall
demand that the variation with respect to the dipole moment provides
an independent evolution equation.

The introduction of the scalar potential function $V$ is motivated by
the quasi Newtonian model of paper~I, consisting of a pair of
sub-particles, with opposite gravitational masses $\pm m$ and positive
inertial masses $+m$ (in analogy with the electric dipole in
electrostatics), and interacting \textit{via} some non-gravitational
force making possible the existence of a stable dipolar configuration.
[Indeed, free gravitational masses with opposite sign will accelerate
apart from each other, and consequently must be bound by an internal
force able to counteract their gravitational repulsion.] In the present
relativistic model we shall also need to invoke such a force ---
supposed to describe some non-gravitational attractive interaction
between the constituents of the dipole. This force will derive from the
potential $V$ in the action.

Let us however stress that we shall not use any model for the elementary
constituents of the dipole, nor shall we invoke the notion of a negative
mass. The formalism follows completely from the action~\eqref{S}, which
contains simply the moment $\pi^\mu$ and a positive mass parameter $m$.
As already commented in the introduction, the mass $2m$ represents the
total inertial mass $M_\mathrm{i}$ of the dipolar particle (in agreement
with paper~I), but here, contrary to paper~I, this mass can also be
interpreted as the passive gravitational mass $M_\mathrm{p}$,
\textit{i.e.} $M_\mathrm{i}=M_\mathrm{p}=2m$. Later we shall argue that
one can also define an active gravitational mass $M_\mathrm{a}$ by the
monopole part of the particle's stress-energy tensor, and that this mass
turns out to be negligible in the NR limit,
$M_\mathrm{a}=\mathcal{O}(c^{-2})$.\,\footnote{Notice the difference
with the quasi-Newtonian ``microscopic'' model of paper~I, in which we
have $M_\mathrm{i}=2m$ and $M_\mathrm{p}=M_\mathrm{a}=0$. This
represents the fundamental difference between the model of paper~I and
the (NR limit of the) present relativistic model.}

\subsection{Equations of motion and evolution}\label{secIIB}
By the principle of stationary action we require that $S$ be unchanged
under the infinitesimal variation $\delta x^\mu(\tau)$ and
$\delta\pi^\mu(\tau)$ in the dynamical variables, with the condition
that they vanish on the boundary of the region of integration,
\textit{i.e.} $\delta x^\mu(\pm\infty)=0$ and
$\delta\pi^\mu(\pm\infty)=0$. We provide in Appendix~\ref{appA} below
a summary of the way we vary the dipolar action.

Varying first $S$ with respect to the change $\delta\pi^\mu$ in the
dipole moment, holding the particle's world-line fixed, $\delta x^\mu
= 0$, we obtain an equation taking the form of the force law
\begin{equation}\label{Pdot}
\dot{P}^\mu = - 2 F^\mu\,,
\end{equation}
where we remind that the dot means the covariant derivative with respect
to proper time: $\dot{P}^\mu\equiv D P^\mu/d\tau$. In the left-hand-side
(LHS) we have defined what will turn out to be the linear momentum
$P^\mu$ of the dipolar particle; it is given in terms of the space-like
vector defined in Eq.~\eqref{pmu} by
\begin{equation}\label{Pmu}
P^\mu = p^\mu + \dot{\pi}^\mu\,.
\end{equation}
As we see $P^\mu$ differs from $p^\mu$, and we shall check later that
$P^\mu$ is time-like. In the right-hand-side (RHS) of~\eqref{Pdot},
$F^\mu$ is the quadri-force derived from the potential $V$ present in
the action, and is given by
\begin{equation}\label{Fmu}
F^\mu = \frac{\pi_\perp^\mu}{\pi_\perp}
\,\frac{dV}{d x}\!\left(\frac{\pi_\perp}{m}\right)\,.
\end{equation}
Here $dV/d x$ is the derivative of $V$ with respect to its natural
argument $x=\pi_\perp/m$. This force is proportional to the orthogonal
projection of the dipole moment~\eqref{piperp}, thus it satisfies the
constraint
\begin{equation}\label{umuFmu}
u_\mu F^\mu=0\,.
\end{equation}

We next perform a variation $\delta x^\mu$ of the particle's position,
holding the components of the dipole moment $\pi^\mu$ ``constant''
during the displacement of the world-line. For instance one can think
of the dipole moment vector as being transported parallely along the
displacement vector $\delta x^\mu$. We find that the variation of the
covariant time derivative of the dipole moment,
$\dot{\pi}^\mu=D\pi^\mu/d\tau$, yields a Riemann curvature term, which
can be understood as coming from the non commutation of the
variational derivative $\delta$ with the covariant derivative
$D/d\tau$. During the variation we must take into account the
contribution due to the orthogonal projection operator $\perp^\mu_\nu$
present in the term $V(\pi_\perp/m)$. All computations done we end up
with the equation (where $\dot{\Omega}^\mu\equiv D\Omega^\mu/d\tau$)
\begin{equation}\label{Omegadot}
\dot{\Omega}^\mu = - \frac{1}{2m} R^\mu_{\,\,\,\rho\nu\sigma} u^\rho
\pi^\nu P^\sigma \,.
\end{equation}
See Appendix~\ref{appA} for a general derivation of this equation.
The RHS represents the analogue of the famous Papapetrou coupling to
the Riemann curvature tensor in the equation of motion of a particle
with spin~\cite{papa51spin}.\,\footnote{In the case of a spinning
particle the antisymmetric spin tensor is given in terms of the dipole
moment variable and the particle's linear momentum by
$S^{\mu\nu}=\pi^{[\mu}P^{\nu]}/m$.} The ``linear momentum''
$\Omega^\mu$ in the LHS of~\eqref{Omegadot} is found to be different
from $P^\mu$ and to be given by
\begin{equation}\label{Om}
\Omega^\mu = \omega^\mu - p^\mu\,,
\end{equation}
in which we used again the convenient definition of $p^\mu$,
Eq.~\eqref{pmu}, and where $\omega^\mu$ represents another intermediate
quantity given by
\begin{equation}\label{om}
\omega^\mu =
\frac{u^\mu}{c^2}\left(\frac{\dot{\pi}_\nu\,\dot{\pi}^\nu}{4m} +
V\right) - \frac{u_\nu\pi^\nu}{m c^2} F^\mu\,.
\end{equation}
At this point we observe that the complete dynamics of the dipolar
particle is encoded into the two equations~\eqref{Pdot}
and~\eqref{Omegadot}.

Next we obtain the particle's stress-energy tensor by varying the
action~\eqref{S} with respect to an infinitesimal change in the
background metric, $\delta g_{\mu\nu}$, vanishing at the edges of the
space-time manifold, when $\vert
x^\mu\vert\rightarrow\infty$. Obviously we must take into account all
metric contributions, including crucially those arising from the
Christoffel symbols in the covariant time derivative $\dot{\pi}^\mu$,
and those coming from $\perp_{\mu\nu}=g_{\mu\nu}+u_\mu u_\nu/c^2$. The
conserved number density $n$ of the dipolar particles satisfies the
covariant continuity equation
\begin{equation}\label{ncons}
\nabla_\nu (n\,u^\nu)=0\,.
\end{equation}
By straightforward calculations --- see Appendix~\ref{appA} --- we
find that the stress-energy tensor $T^{\mu\nu}$ (with the dimension of
an energy density) of the dipolar particles can be expressed in terms
of the two basic linear momenta $\Omega^\mu$ and $P^\mu$ as
\begin{equation}\label{Tmunu}
T^{\mu\nu} = n \,\Omega^{(\mu} u^{\nu)} - \frac{1}{2m}
\nabla_\rho\Bigl(n \Bigl[\pi^\rho P^{(\mu} - P^\rho
\pi^{(\mu}\Bigr]u^{\nu)}\Bigr) \,.
\end{equation}
We readily verify that the covariant conservation law
\begin{equation}\label{divTmunu}
\nabla_\nu T^{\mu\nu}=0\,,
\end{equation}
holds as a consequence of the equations of motion~\eqref{Pdot}
and~\eqref{Omegadot}.

Some physical interpretation follows from the expression of
$T^{\mu\nu}$. It is clear that the first term in~\eqref{Tmunu} takes
the form of a \textit{monopolar} contribution, appropriate for a
point-like particle having velocity $u^\mu$ and linear momentum
$\Omega^\mu$. We thus see that the particles we are considering are
not purely dipolar, since they involve also a monopolar contribution
in this sense. The monopolar piece in the stress-energy tensor will
generate a monopolar gravitational field \textit{via} the Einstein
field equations. So the mass associated with the linear momentum
$\Omega^\mu$ can be naturally interpreted --- since it represents the
point-like source of a monopolar gravitational field --- as the active
gravitational mass of the particle [see~\eqref{Ma} below for the
computation of this mass within a particular solution of the
equations]. Similarly $\Omega^\mu$ can be referred to as the ``active
linear momentum'', while $P^\mu$ which enters the law of
motion~\eqref{Pdot} should rather be regarded as the ``inertial linear
momentum''. [Technically $\Omega_\mu$ represents the conjugate
momentum of $x^\mu$ and $P_\mu$ the conjugate momentum of $\pi^\mu$;
see Eqs.~\eqref{conjPOm} in the Appendix.]

The second term in Eq.~\eqref{Tmunu} is clearly dipolar, and represents
the relativistic generalization of the quasi-Newtonian density of
polarization $\rho_\mathrm{polar}=-\partial_i\Pi^i$ in paper~I ---
\textit{i.e.} minus the divergence of the polarization vector $\Pi^i$.
To emphasize the dipolar character of this term we rewrite the
stress-energy tensor as
\begin{equation}\label{Tmunu'}
T^{\mu\nu} = n \,\Omega^{(\mu} u^{\nu)} -
c^2\nabla_\rho\Pi^{\rho\mu\nu}\,,
\end{equation}
where $\Pi^{\rho\mu\nu}$ can be called the polarization tensor and is
given by
\begin{equation}\label{Pol}
\Pi^{\rho\mu\nu} = \frac{n}{2m c^2}\Bigl[\pi^\rho P^{(\mu} - P^\rho
\pi^{(\mu}\Bigr]u^{\nu)}\,.
\end{equation}
The gravitational field generated by the distribution of dipolar
particles will be computed in the standard way by adding the matter
action~\eqref{S} to the Einstein-Hilbert action for the gravitational
field. Equivalently we shall put the stress-energy
tensor~\eqref{Tmunu} in the RHS of the Einstein field equations of
general relativity.  (Obviously we can use also the gravitational
action and field equations of any favorite metric theory of gravity.)

Finally let us exploit some constraint equations which are satisfied
by the general solution of the equations~\eqref{Pdot}
and~\eqref{Omegadot}. First we notice that a consequence of
Eq.~\eqref{Omegadot}, implied by the antisymmetry of the Riemann
tensor with respect to its first pair of indices, is
\begin{equation}\label{uOmdot}
u_\mu\dot{\Omega}^\mu=0\,.
\end{equation}
Let us prove that in fact this relation is identically satisfied, in the
sense that it is implied by the other equations we have. We recall that
$\Omega^\mu=\omega^\mu-p^\mu$. Using the definition~\eqref{om} we
compute first the contraction $u_\mu\dot{\omega}^\mu$, and reduce it
thanks to the easy-to-check formula
\begin{equation}\label{Vdot}
\dot{V} = \frac{1}{m}F_\mu \dot{\pi}_\perp^\mu\,,
\end{equation}
which follows from the fact that $V$ is a function of $\pi_\perp/m$
only. In Eq.~\eqref{Vdot} we have used the expression of the
quadri-force~\eqref{Fmu} and we have defined\,\footnote{Beware of the
fact that since the motion of the particle will not be geodesic
($\dot{u}^\mu\not= 0$), the latter definition $\dot{\pi}_\perp^\mu
\equiv D[\perp^\mu_\nu \pi^\nu]/d\tau$ is in general \textit{different}
from $\perp^\mu_\nu D\pi^\nu/d\tau$. Thus one is not allowed to commute
the operations of perpendicular projection $\perp^\mu_\nu$ and of taking
the covariant time derivative $D/d\tau$.}
\begin{equation}\label{dotpiperp}
\dot{\pi}_\perp^\mu \equiv \frac{D\pi_\perp^\mu}{d\tau}\,.
\end{equation}
At this stage we arrive at the intermediate result $u_\mu
\dot{\omega}^\mu=-\dot{\pi}_\mu(\ddot{\pi}^\mu+2F^\mu)/(2m)$. Next we
have recourse to the other equation~\eqref{Pdot}, and we remind the
fact that $p_\mu\dot{p}^\mu=0$ since the norm of $p^\mu$ is a
constant, $p^2=4m^2c^2$. This yields immediately $u_\mu
\dot{\omega}^\mu=u_\mu \dot{p}^\mu$ which is indeed the
constraint~\eqref{uOmdot} we wanted to prove. So this constraint is
consistent with our basic equations~\eqref{Pdot} and~\eqref{Omegadot}.

We now show that there is another constraint relation, which is a
non-trivial consequence of our equations, and will be used below to
find an interesting particular solution of those equations. This
relation is obtained by contracting Eq.~\eqref{Pdot} with the four
velocity. Because $F^\mu$ is orthogonal to the four-velocity
[\textit{cf} Eq.~\eqref{umuFmu}], we obtain 
\begin{equation}\label{uPdot}
u_\mu\dot{P}^\mu=0\,,
\end{equation}
which can easily be transformed, using the definition~\eqref{Pmu} of
$P^\mu$, into\,\footnote{We can write also the alternative equivalent
expression
$$\dot{\Lambda}(2m+u_\mu\dot{\pi}^\mu)+\Lambda(\Lambda-1)u_\mu\ddot{\pi}^\mu=0\,.$$}
\begin{equation}\label{uPexpl}
\dot{\Lambda}\,u_\mu p^\mu+(\Lambda-1)\,u_\mu\dot{p}^\mu=0\,.
\end{equation}
The constraint relation~\eqref{uPexpl} could be viewed as a differential
equation for the quantity $\Lambda$ which is defined by~\eqref{Lambda}.
Since it is a consequence of our main equations~\eqref{Pdot}
and~\eqref{Omegadot}, the relation~\eqref{uPexpl} is to be satisfied by
any solutions of those equations.

\subsection{Particular solution of the equations}\label{secIIC}
The general dynamics of dipolar particles in the present approach is
given by Eqs.~\eqref{Pdot} and~\eqref{Omegadot}, whose consequence as we
have just seen is Eq.~\eqref{uPexpl}. From now on we shall restrict our
attention to a particular class of solutions owning some features that
are consistent, as we shall see, with the intuitive idea of a dipole
moment in ``equilibrium'' in the gravitational field. This class of
solutions is obtained by solving the constraint equation~\eqref{uPexpl}
in the simplest way that
\begin{equation}\label{Lambda1}
\Lambda = 1\,.
\end{equation}
This choice will greatly simplify the equations of motion~\eqref{Pdot}
and~\eqref{Omegadot}, and make them quite attractive. We shall however
leave open the question of how generic the solutions
satisfying~\eqref{Lambda1} are, and under which conditions (if any)
could they be made unique.\,\footnote{One could be tempted to replace
$\Lambda=1$ back into the original action~\eqref{S}, therefore defining
the alternative action
$$\widetilde{S} = \int d\tau \left[-2m\,c^2 +
\frac{\dot{\pi}_\mu\,\dot{\pi}^\mu}{4m} - V\right]\,.$$ However this
action would not describe a dipolar particle in the sense we want.
Notably since as we see $\widetilde{S}$ involves now a mass term in the
ordinary sense (with mass $2m$), its stress-energy tensor will contain
some unwanted ``monopolar'' mass contribution $\sim
m\,n\,u^\mu\,u^\nu$.} In more details the relation~\eqref{Lambda1} reads
\begin{equation}\label{Lambda1expl}
\frac{\dot{\pi}_\mu\,\dot{\pi}^\mu}{4m} - u_\mu \dot{\pi}^\mu =
2m\,c^2\,.
\end{equation}
This can be regarded as an equation giving the time-like component of
$\dot{\pi}^\mu$, which is parallel to the velocity $u^\mu$
(\textit{i.e.}  $u_\mu\dot{\pi}^\mu$), in terms of the space-like
components, perpendicular to $u^\mu$. The solution
of~\eqref{Lambda1expl} is best expressed in terms of the vector
$p^\mu$ (which is now $p^\mu=2m\,u^\mu-\dot{\pi}^\mu$ because
$\Lambda=1$), and we get
\begin{equation}\label{Lambda1sol}
u_\mu p^\mu/c =
\varepsilon\sqrt{\perp_{\mu\nu}\!\dot{\pi}^\mu\dot{\pi}^\nu - 4
m^2\,c^2}\,,
\end{equation}
where $\varepsilon=\pm 1$ tells whether $p^\mu$ is future or past directed.

Using Eq.~\eqref{Lambda1} it will become clear that the linear
momentum $P^\mu$ represents the flow of inertial or equivalently
passive-type gravitational mass, while the other linear momentum
$\Omega^\mu$ is associated with some active-type gravitational
mass. Furthermore, for the class of solutions
satisfying~\eqref{Lambda1}, it will happen most remarkably that the
``physical'' dipole moment, namely the one which appears in the final
equations of motion and stress-energy tensor, is the projection
orthogonal $\pi_\perp^\mu$ to the four velocity, rather than $\pi^\mu$
itself. Interestingly, we shall find that the longitudinal component,
$u_\mu\pi^\mu$, which never appears in the final equations and is
therefore unphysical (\textit{i.e.} unobservable), is actually given
by a complex number, see Eq.~\eqref{complexupi} below.

From Eqs.~\eqref{pmu} and~\eqref{Pmu} we find that when $\Lambda=1$
the linear momentum $P^\mu$ is simplified to the following vector,
which is time-like,
\begin{equation}\label{Psimple}
P^\mu = 2m\,u^\mu\,.
\end{equation}
Then~\eqref{uPdot} is satisfied simply because the quadri-norm of
$u^\mu$ is constant. Thus the equation of motion~\eqref{Pdot} gives
the particle's quadri-acceleration $a^\mu\equiv\dot{u}^\mu$ as
\begin{equation}\label{mamu}
2m\,a^\mu = -2F^\mu\,.
\end{equation}
Since the mass coefficient $2m$ is in factor of the acceleration (which
incorporates both inertial and gravitational effects), it can
equivalently be interpreted as the particle's inertial mass and passive
gravitational mass, $M_\mathrm{i}=M_\mathrm{p}=2m$; in this sense the
equivalence principle is satisfied. However the motion is not geodesic
as the result of the force $F^\mu$ which is supposed to reflect the
internal structure of the dipolar particle.

Let us reduce next the expression of $\Omega^\mu$, defined by~\eqref{Om}
and~\eqref{om}. To this end we make use of Eq.~\eqref{Lambda1expl} to
obtain first the alternative expression
\begin{equation}\label{Omsimple0}
\Omega^\mu = \frac{u^\mu}{c^2}\bigl[V + u_\nu \dot{\pi}^\nu\bigr] +
  \dot{\pi}^\mu - \frac{u_\nu\pi^\nu}{m c^2} F^\mu\,.
\end{equation}
The point is that we can express $\Omega^\mu$ entirely in terms of the
orthogonal projection $\pi_\perp^\mu = \perp^\mu_\nu\pi^\nu$. We
replace $\dot{\pi}^\mu$ in~\eqref{Omsimple0} by its equivalent
expression in terms of the time derivative of $\pi_\perp^\mu$, namely
$\dot{\pi}_\perp^\mu=D\pi_\perp^\mu/d\tau$ already defined
in~\eqref{dotpiperp}, and which we recall is different from the
alternative object $(\dot{\pi}^\mu)_\perp=\perp^\mu_\nu\dot{\pi}^\mu$.
An easy computation, using also Eq.~\eqref{mamu}, brings then
$\Omega^\mu$ into the simple form
\begin{equation}\label{Omsimple}
\Omega^\mu = \frac{V}{c^2}\,u^\mu \,+
\perp^\mu_\nu\dot{\pi}_\perp^\nu\,,
\end{equation}
displaying the longitudinal \textit{vs} perpendicular decomposition of
$\Omega^\mu$ with respect to the four velocity. Notice that in the
second term of~\eqref{Omsimple} the orthogonal projector $\perp^\mu_\nu$
appears two times: one explicitly in front of the term and one contained
into $\pi_\perp^\nu$. The expression~\eqref{Omsimple} makes it clear
that
\begin{equation}\label{uOm}
u_\mu \Omega^\mu + V = 0\,.
\end{equation}

Motivated by the fact that $\Omega^\mu$ parametrizes the ``mass term''
or ``monopole part'' of the stress-energy tensor~\eqref{Tmunu}, we
define the particle's \textit{active} gravitational mass $M_\mathrm{a}$
--- namely the mass which ``actively'' generates the gravitational field
--- as the coefficient of the velocity $u^\mu$ in Eq.~\eqref{Omsimple}.
More precisely, $M_\mathrm{a}$ is defined by the longitudinal part of
$\Omega^\mu$ along $u^\mu$ as\,\footnote{Such definition is proposed
here for heuristic discussion, but is not used directly in the
formalism.}
\begin{equation}\label{Ma}
M_\mathrm{a} \equiv -\frac{1}{c^2}\,u_\mu \Omega^\mu = \frac{V}{c^2} \,.
\end{equation}
This mass is not conserved because of the work done by the internal
force $F^\mu$, and we find [consistently with Eq.~\eqref{Vdot}]
\begin{equation}\label{Madot}
\dot{M}_\mathrm{a} = \frac{1}{m c^2}\,F_\mu \Omega^\mu\,.
\end{equation}
With this notation the equation~\eqref{Omsimple} can be viewed as the
classic relation between the particle's linear momentum $\Omega^\mu$
and the four velocity $u^\mu$, namely $\Omega^\mu =
M_\mathrm{a}\,u^\mu + \text{(dipolar effects)}$. A similar relation
holds in the case of spinning particles, where the linear momentum and
velocity differ by spin effects~\cite{papa51spin, Traut58, Dixon}.

The evolution equation~\eqref{Omegadot} reads now\,\footnote{In more
details this equation can also be written as [making use of
Eqs.~\eqref{Vdot} and~\eqref{mamu}]
$$\Bigl(V + u_\nu\dot{\pi}_\perp^\nu\Bigr)\frac{a^\mu}{c^2}\, +
\perp^\mu_\nu\ddot{\pi}_\perp^\nu = - \pi_\perp^\nu
\,R^\mu_{\,\,\,\rho\nu\sigma} u^\rho u^\sigma \,.$$}
\begin{equation}\label{Omdots}
\dot{\Omega}^\mu \equiv \frac{D}{d\tau}\biggl[\frac{V}{c^2}\,u^\mu +
\perp^\mu_\nu\dot{\pi}_\perp^\nu\biggr] = - \pi_\perp^\nu
\,R^\mu_{\,\,\,\rho\nu\sigma} u^\rho u^\sigma \,,
\end{equation}
where we have taken advantage of the symmetries of the Riemann tensor to
replace $\pi^\nu$ by $\pi_\perp^\nu$ in the RHS. This point is not
without interest because we discover that the equation depends only on
the orthogonal projection $\pi_\perp^\mu$ --- indeed recall that $V$
itself is a function of
$\pi_\perp=\sqrt{g_{\mu\nu}\pi^\mu_\perp\pi^\nu_\perp}$. In
Sect.~\ref{secIIIA} we shall interpret the non relativistic limit of
Eq.~\eqref{Omdots} as an equilibrium condition for the dipole moment in
a gravitational field.

Finally the stress-energy tensor for the class of solutions satisfying
$\Lambda=1$ is given by
\begin{equation}\label{Tmunubis}
T^{\mu\nu} = n \biggl[\frac{V}{c^2}\,u^{\mu} u^{\nu} +
u^{(\mu}\!\perp^{\nu)}_\rho\dot{\pi}_\perp^\rho\biggr] -
c^2\nabla_\rho\Pi^{\rho\mu\nu}\,,
\end{equation}
where the polarization tensor~\eqref{Pol} reads now
\begin{equation}\label{polarbis}
\Pi^{\rho\mu\nu} = \frac{n}{c^2} \,\Bigl[\pi_\perp^\rho u^{(\mu} -
u^\rho \pi_\perp^{(\mu}\Bigr]u^{\nu)} \,.
\end{equation}
Here also we have been able to replace $\pi^\mu$ by $\pi_\perp^\mu$,
thanks to the antisymmetry of the two terms in~\eqref{polarbis}.

To conclude, the choice of solution~\eqref{Lambda1} enables one to
appreciably simplify the equations and to ease their interpretation. A
consequence is that the number of independent components of the dipole
moment is reduced from four down to three. The component of the dipole
moment that is along the four velocity (namely $u_\mu\pi^\mu$) never
appears in the final equations and is unobservable. The physical
dipole moment is entirely described by the orthogonal projection
variable $\pi_\perp^\mu$, which is a \textit{space-like} vector.

\section{Dipolar particles in the non relativistic limit}\label{secIII}
\subsection{Quasi Newtonian equations}\label{secIIIA}
We investigate the non-relativistic (NR) approximation of the dynamics
of dipolar particles, as described by the solution found in
Sect.~\ref{secIIC}, characterized by the equations of
motion~\eqref{mamu} and~\eqref{Omdots}, and by the stress-energy
tensor~\eqref{Tmunubis}--\eqref{polarbis}. To proceed, we write
explicitly all factors of $c$'s, and consider the formal limit when
$c\rightarrow\infty$, which is equivalent to the usual $v/c\rightarrow
0$, where $v$ is the typical value of the coordinate velocity of the
dipolar particles. In the following we systematically indicate the
neglected remainder terms and write them as some $\mathcal{O}(c^{-n})$.

We suppose that the dipolar particles evolve in the gravitational field
of some ordinary matter system with spatially compact support. We
introduce a Cartesian coordinate grid $\{t,x^i\}$ (with $t=x^0/c$), and
we choose it to be inertial, \textit{i.e.} without rotation nor
acceleration with respect to some averaged cosmological matter
distribution at large distances from the local matter system. In these
coordinates the metric is asymptotically flat (the local matter system
is freely falling in the cosmological background field).

In the NR limit the gravitational field is described by a single
Newtonian-like potential $U$, whose source will be the sum of the
Newtonian densities of the ordinary matter and of the dipolar particles.
Such potential will satisfy a Poisson equation coming from the NR limit
of the Einstein field equations. In the usual notation $U$ has the
dimension of a velocity squared and is of order one when
$c\rightarrow\infty$, which we denote by $U=\mathcal{O}(c^0)$. The
metric coefficients are given by
\begin{subequations}\label{metric}\begin{align}
g_{00} &= -1+\frac{2U}{c^2}+\mathcal{O}\left(\frac{1}{c^4}\right)\,,\\
g_{0i} &= \mathcal{O}\left(\frac{1}{c^3}\right)\,,\\ g_{ij} &=
\delta_{ij}\left(1+\frac{2U}{c^2}\right)+\mathcal{O}\left(\frac{1}{c^4}\right)\,.
\end{align}\end{subequations}
This metric is accurate enough to obtain the Poisson equation satisfied
by $U$ in the NR approximation, and to discuss the motion of massive
particles (ordinary stars) as well as relativistic particles (ordinary
photons) in the gravitational field.

The four-velocity $u^\mu$ of the dipolar particle is written as
$u^\mu=(u^0,u^0v^i/c)$, where $v^i=d x^i/d t$ is the coordinate velocity
and $u^0=c \,d t/d\tau$. Since the particle will be non relativistic,
$v^i$ is of order unity when $c\rightarrow\infty$, \textit{i.e.}
$v^i=\mathcal{O}(c^0)$. In the NR limit the proper time reduces to the
coordinate time, $d\tau = d t + \mathcal{O}(c^{-2})$, hence $u^0 = c +
\mathcal{O}(c^{-1})$ and $u^i = v^i + \mathcal{O}(c^{-2})$. The
particle's quadri-acceleration, given by $a^\mu=d u^\mu/d\tau +
\Gamma^\mu_{\rho\sigma}u^\rho u^\sigma$, is then found to reduce in the
NR limit to $a^0 = \mathcal{O}(c^{-1})$ and $a^i = d^2 x^i/d t^2 + c^2
\Gamma^i_{00} + \mathcal{O}(c^{-2})$, where $d^2 x^i/d t^2$ is the
ordinary coordinate acceleration. The Christoffel symbol reads
$\Gamma^i_{00}=-c^{-2}g^i + \mathcal{O}(c^{-4})$, where $g^i =
\partial_i U$ is the Newtonian gravitational field, so we have
\begin{equation}\label{ai}
a^i = \frac{d^2 x^i}{d t^2} - g^i +
\mathcal{O}\left(\frac{1}{c^2}\right)\,.
\end{equation}
[Of course, geodesic motion simply means that $a^i=0$ hence $d^2 x^i/d
t^2=g^i + \mathcal{O}(c^{-2})$.]

Consider the internal force $F^\mu$, which is defined by Eq.~\eqref{Fmu}
and appears in the RHS of the law of motion~\eqref{mamu}. As a basic
(and quite natural) hypothesis, we impose that this force exists in the
NR limit, in the sense that its \textit{spatial} components $F^i$ admit
a non-zero and finite limit when $c\rightarrow\infty$:
\begin{equation}\label{Fi}
F^i = \mathcal{O}\left(c^0\right)\,.
\end{equation}
Since $F^\mu$ is orthogonal to the velocity, Eq.~\eqref{umuFmu}, its
zero-th component $F^0$ vanishes in the NR limit, $F^0 =
\mathcal{O}(c^{-1})$ [and more precisely we have $F^0 = v^i F^i/c +
\mathcal{O}(c^{-3})$]. It is now clear that the dipolar particle's law
of motion~\eqref{mamu} reduces in the NR approximation to the
non-geodesic equation\,\footnote{Since both $a^0$ and $F^0$ are of order
$\mathcal{O}(c^{-1})$ we see that the zero-th component of the law of
motion is also satisfied in the NR limit.}
\begin{equation}\label{eom}
m\,\frac{d^2 x^i}{d t^2} = m\,g^i - F^i +
\mathcal{O}\left(\frac{1}{c^2}\right)\,.
\end{equation}
We emphasize that $g^i = \partial_i U$ represents here the
\textit{total} gravitational field, which is sourced not only by the
ordinary matter but also by the dipolar particles themselves.

Consistently with the order of magnitude~\eqref{Fi} we assume that the
potential function $V$ from which derives the force $F^i$ is also of the
same order,
\begin{equation}\label{V0}
V = \mathcal{O}\left(c^0\right)\,.
\end{equation}
Equation~\eqref{V0} means that we preclude any constant term in the
expression of $V$, which would be of the order of $\mathcal{O}(c^2)$
and therefore be of the form $\mathcal{M}\,c^2$, where $\mathcal{M}$
is some constant mass parameter. Such term corresponds to a mass term
in the action~\eqref{S}, and would imply that the dipolar particle is
endowed not only by the dipole moment $\pi^\mu$ (and its associated
mass $m$), but also by the mass $\mathcal{M}$ in the ordinary sense
--- dark matter particles would carry some mass in the ordinary
sense. This $\mathcal{M}$ yields an unwanted monopolar contribution to
the stress-energy tensor (non vanishing in the NR limit), so we simply
pose $\mathcal{M}=0$. Related to this we note that Eq.~\eqref{V0}
implies that the active gravitational mass $M_\mathrm{a}=V/c^2$ we
defined in~\eqref{Ma} vanishes in the NR limit:
\begin{equation}\label{Ma0}
M_\mathrm{a} = \mathcal{O}\left(\frac{1}{c^2}\right)\,.
\end{equation}
In the NR limit the dipolar particle has zero active-type
gravitational mass, hence its stress-energy tensor is purely dipolar,
in agreement with the elementary intuition from paper~I.

Consider next the other equation~\eqref{Omdots}. Our basic dipole moment
variable $\pi_\perp^\mu$ is assumed to be such that [consistently
with~\eqref{Fi} and~\eqref{V0}]
\begin{equation}\label{pii0}
\pi_\perp^i = \mathcal{O}\left(c^0\right)\,,
\end{equation}
hence $\pi_\perp^0=\mathcal{O}(c^{-1})$ which follows from the
orthogonality to the velocity. The fact that $\pi_\perp^0$ vanishes in
the NR limit is crucial. It implies that the four norm of
$\pi_\perp^\mu$ reduces to the Euclidean norm of $\pi_\perp^i$ in the
NR limit:
$\pi_\perp=[\delta_{ij}\pi_\perp^i\pi_\perp^j]^{1/2}+\mathcal{O}(c^{-2})$.
Hence the force $F^i$ is a function only of $\pi_\perp^j$ in this
limit.\,\footnote{More precisely we should write that
$F^i[\pi_\perp^\mu]=\tilde{F}^i[\pi_\perp^j]+\mathcal{O}(c^{-2})$, but
in the following we shall identify, by a slight abuse of notation, the
functional $\tilde{F}^i$ with the original one $F^i$.} We need the
covariant proper time derivative of $\pi_\perp^\mu$ which is given by
$\dot{\pi}_\perp^\mu=d \pi^\mu_\perp/d\tau +
\Gamma^\mu_{\nu\rho}u^\nu\pi_\perp^\rho$. Because $\pi_\perp^0$
vanishes in the NR limit, we readily find that $\dot{\pi}_\perp^\mu$
is not coupled to gravity in this limit, in the sense that the
Christoffel symbols make a contribution which is of higher
order. Hence we find that $\dot{\pi}_\perp^0=\mathcal{O}(c^{-1})$, and
most importantly that the spatial components $\dot{\pi}_\perp^i$
reduce to the \textit{ordinary} time derivative, namely
$\dot{\pi}_\perp^i=d \pi^i_\perp/d t + \mathcal{O}(c^{-2})$. The
linear momentum~\eqref{Omsimple} then becomes
$\Omega^0=\mathcal{O}(c^{-1})$ and $\Omega^i=d \pi^i_\perp/d t +
\mathcal{O}(c^{-2})$. Applying the same reasoning we find that
$\dot{\Omega}^0=\mathcal{O}(c^{-1})$ and $\dot{\Omega}^i=d^2
\pi^i_\perp/d t^2 + \mathcal{O}(c^{-2})$. Finally the Riemann
curvature tensor in the RHS of~\eqref{Omdots} yields a coupling to the
tidal gravitational field $\partial_{ij} U$. Therefore we have proved
that the NR limit of this equation is
\begin{equation}\label{eveq}
\frac{d^2 \pi_\perp^i}{d t^2} = \pi_\perp^j \partial_{ij} U +
\mathcal{O}\left(\frac{1}{c^2}\right)\,.
\end{equation}
The law of evolution given by Eq.~\eqref{eveq} is interesting for our purpose
because it can be regarded as a condition of equilibrium for the dipole moment
in a gravitational field. It states that when the tidal gravitational field
can be neglected, the components of the dipole moment $\pi_\perp^i$ stay
constant (or evolve linearly in time). Note that Eq.~\eqref{eveq} is formally
identical with the equation of \textit{geodesic deviation} for the
(space-like) ``separation'' vector $\pi_\perp^\mu$ in the NR limit. However,
because of other terms in Eq.~\eqref{Omdots}, the equation of geodesic
deviation will not hold outside this limit.

Thus our equations are the law of motion~\eqref{eom} and the
equilibrium condition~\eqref{eveq}. Let us emphasize that these
equations have a structure different from those of paper~I, which were
concocted by analogy with the model of a dipole moment in
electrostatics, and in particular violate the equivalence
principle. In the present relativistic description of the dipole, we
are consistent with the equivalence principle, and as a result the law
of motion~\eqref{eom} is different from Eq.~(12) in paper~I. We also
obtain directly an equilibrium condition~\eqref{eveq} for the dipole
moment, instead of the evolution equation (13) of paper~I. Amazingly,
we find that in order to reproduce the present equations~\eqref{eom}
and~\eqref{eveq} of the (NR limit of the) relativistic model, the two
RHS of Eqs.~(12) and~(13) in paper~I should exactly be interchanged.

As we pointed out, because of the constraint relation~\eqref{Lambda1},
the end equations depend only on the orthogonal projection of the
dipole moment $\pi_\perp^\mu$ (or $\pi_\perp^i$ in the NR
limit). Still it is interesting to control the original, unprojected
dipole moment variable $\pi^\mu$ which entered into the
action~\eqref{S}. We do it here, by considering Eq.~\eqref{Lambda1sol}
in the NR limit. From the results $\dot{\pi}^0=\mathcal{O}(c)$ and
$\dot{\pi}^i=\mathcal{O}(c^0)$, we find that
$\perp_{\mu\nu}\!\dot{\pi}_\perp^\mu\dot{\pi}_\perp^\nu=\mathcal{O}(c^0)$
(because of the orthogonal projection), hence this term is negligible
with respect to the mass term in Eq.~\eqref{Lambda1sol}. This equation
thus becomes $u_\mu
p^\mu=2\varepsilon\,\mathrm{i}\,m\,c^2+\mathcal{O}(c^0)$ where
$\mathrm{i}=\sqrt{-1}$ is the imaginary number (and $\varepsilon=\pm
1$). The appearance of a complex quantity is due to the space-like
character of $p^\mu$ (which satisfies $p^2=4m^2\,c^2$). Now we further
deduce $p^0=-2\varepsilon\,\mathrm{i}\,m\,c + \mathcal{O}(c^{-1})$ and
$p^i=\mathcal{O}(c^0)$ in the NR limit. The interesting finding is
that the components of $\pi^\mu$ are seen to be \textit{complex} in
the NR approximation for that solution, and given by
\begin{subequations}\label{complex}\begin{align}
\pi^0 &= 2m\,c\left(1+\varepsilon\,\mathrm{i}\right)t +
\mathcal{O}\left(\frac{1}{c}\right)\,,\\ \pi^i &= \pi_\perp^i +
2m\,v^i\left(1+\varepsilon\,\mathrm{i}\right)t +
\mathcal{O}\left(\frac{1}{c^2}\right)\,.
\end{align}\end{subequations}
In particular the longitudinal component along the velocity is
\begin{equation}\label{complexupi}
u_\mu \pi^\mu = -2m\,c^2\left(1+\varepsilon\,\mathrm{i}\right)t +
\mathcal{O}\left(c^0\right)\,.
\end{equation}
However, the components of $\pi^\mu$ itself do not appear into the
final equations of motion and stress-energy tensor, and are therefore
unphysical; only the components of $\pi_\perp^\mu$, which parametrize
the final equations, represent the physical variables, and these are
(to be considered as) real.

We compute the stress-energy tensor~\eqref{Tmunubis}--\eqref{polarbis}
in the NR limit. In such limit the number density $n$ of dipolar
particles satisfies the Eulerian continuity equation,
\begin{equation}\label{continuity}
\partial_t \,n + \partial_i \bigl(n \,v^i\bigr) =
\mathcal{O}\left(\frac{1}{c^2}\right)\,.
\end{equation}
From Eq.~\eqref{V0} [or equivalently~\eqref{Ma0}], together with the
fact that $\dot{\pi}^0_\perp = \mathcal{O}(c^{-1})$, we deduce that
the monopolar term in Eq.~\eqref{Tmunubis} is zero in the NR limit so
that (most satisfactorily) the stress-energy tensor becomes purely
dipolar. The covariant derivative of the polarization tensor can be
approximated by an ordinary derivative, so the components of
$T^{\mu\nu}$ are simply obtained as
\begin{subequations}\label{TNR}\begin{align}
T^{00} &= - c^2 \partial_i\Pi^i + \mathcal{O}(c^0)\,,\\T^{0i} &=
\mathcal{O}(c)\,,\\T^{ij} &= \mathcal{O}(c^0)\,,
\end{align}\end{subequations}
in which the density of dipole moment or polarization vector is defined
by
\begin{equation}\label{polar}
\Pi^i = n\,\pi_\perp^i\,.
\end{equation}

It is clear from Eqs.~\eqref{TNR} that since the $T^{\mu\nu}$ of dipolar
particles is to be added to the one of the other matter fields, the only
effect of the dipolar particles in the NR limit is to add to the
Newtonian density of the ordinary matter the density of polarization
\begin{equation}\label{rhopol}
\rho_\mathrm{polar}=-\partial_i\Pi^i\,.
\end{equation}
The Einstein field equations reduce in the NR limit to the Poisson
equation for the potential $U$ defined by the metric
coefficients~\eqref{metric}. The point is that the source of the
Poisson equation is the \textit{total} matter density; therefore we
have proved that in the NR limit
\begin{equation}\label{poisson}
\Delta U = -4\pi G\bigl(\rho+\rho_\mathrm{polar}\bigr) +
\mathcal{O}\left(\frac{1}{c^2}\right)\,,
\end{equation}
where $\rho$ denotes the Newtonian density of the ordinary (monopolar)
matter while $\rho_\mathrm{polar}$ is the dipolar matter density found
in~\eqref{rhopol}.

The gravitational field, given by the metric~\eqref{metric} where $U$
is solution of~\eqref{poisson}, affects the dynamics of any matter
distribution. In the case of the dipolar particles we have already
derived the equation of motion in Eq.~\eqref{eom} above. For non
relativistic ordinary particles (``ordinary stars'' assimilated as
point masses) we have the standard acceleration law, coming from the
universal coupling to gravity and the geodesic equation, namely
\begin{equation}\label{eomord}
\Bigl(\frac{d^2 x^i}{d t^2}\Bigr)_\mathrm{ordinary} = g^i +
\mathcal{O}\left(\frac{1}{c^2}\right)\,.
\end{equation}
In the case of the motion of \textit{relativistic} ordinary particles
(photons), we find that the equation of motion is exactly given by what
general relativity predicts, \textit{i.e.}
\begin{equation}\label{eomph}
\Bigl(\frac{d^2 x^i}{d t^2}\Bigr)_\mathrm{photon} =
\bigl[1+\beta^j\beta^j\bigr]\,g^i - 4 \beta^i \beta^j g^j +
\mathcal{O}\left(\frac{1}{c}\right)\,,
\end{equation}
where the velocity of the photon $c\,\beta^i\equiv (d x^i/d
t)_\mathrm{photon}$ satisfies $\beta^j\beta^j \left(1 + 2U/c^2\right)
= 1 - 2U/c^2$. Indeed the result~\eqref{eomph} is the consequence of
the metric~\eqref{metric}\,\footnote{We assumed that the spatial
metric $g_{ij}$ is the general relativistic one, with post-Newtonian
parameter $\gamma=1$. Obviously the calculation could be done for a
general value of the parameter $\gamma$.} and the expression of the
stress-energy tensor~\eqref{TNR}. However recall that $g^i=\partial_i
U$ where $U$ is solution of Eq.~\eqref{poisson}, so the light
deflection of photons is given by the usual formula~\eqref{eomph}, but
in which $U$ is generated by the ordinary stars \textit{and} also by
the distribution of dipolar particles themselves.

The latter point on light deflection constitutes an attractive feature
of the present model of dipolar dark matter. Indeed the distribution
of dark matter is felt not only by ordinary stars (and gas) moving
around galaxies, but also by photons, as is evidenced in experiments
probing the large scale structure using light deflection and
amplification (weak lensing experiments~\cite{Fort}). The problem of
light deflection turned out to be crucial in the construction of
relativistic MOND theories based on extra fields besides the metric
tensor of general relativity~\cite{Sand97, Bek04, Sand05}. To account
for the observed strong light bending in lensing experiments, one is
obliged in these theories to modify the standard conformal coupling of
a scalar field to matter by means of some vector field (preferred or
dynamical) especially designed for this purpose~\cite{Sand97, Bek04,
Sand05}. The resulting tensor-scalar-vector theory is inevitably
complicated. In the present model, by contrast, we have derived in a
natural way the formula for the light bending~\eqref{eomph}, which
takes into account, \textit{via} the Poisson equation~\eqref{poisson},
the effect due to the distribution of dipolar dark matter.

To summarize this Section, the equations giving the dynamics of the
fluid of dipolar particles in the NR limit are:
\begin{itemize}
\item[(i)] Their equation of motion~\eqref{eom};
\item[(ii)] An ``equilibrium'' condition for the dipole
moments~\eqref{eveq};
\item[(iii)] The conservation of the number of particles or continuity
equation~\eqref{continuity};
\item[(iv)] The Poisson equation~\eqref{poisson} for the gravitational
field.
\end{itemize}
In addition we have Eqs.~\eqref{eomord} and~\eqref{eomph} for the motion
of ordinary matter and photons respectively.

\subsection{Link with the modified Newtonian dynamics}\label{secIIIB}
Having arrived at the Poisson equation~\eqref{poisson}, whose source
contains the density of polarization $\rho_\mathrm{polar}$ given by
Eq.~\eqref{rhopol}, we apply the arguments of paper~I and recover the
MOND equation~\cite{Milg1, Milg2, Milg3} when the polarization vector
$\Pi^i$ is aligned with the gravitational field. We pose
\begin{equation}\label{Pigi}
\Pi^i = - \frac{\chi}{4\pi G} \,g^i\,,
\end{equation}
and interpret $\chi$, following paper~I, as a coefficient of
``gravitational susceptibility'' reflecting the properties of the
dipolar medium. We assume that $\chi$ depends on the norm $g=\vert
g^i\vert$ of the gravitational field, in analogy with the electric
susceptibility of a dielectric material which depends on the norm of
the electric field. Next we define
\begin{equation}\label{mu}
\mu = 1 + \chi\,.
\end{equation}
Without loss of generality, we can scale $g$ by means of some constant
acceleration $a_0$, so that~\eqref{mu} defines a certain function
$\mu(g/a_0)$. The Poisson equation~\eqref{poisson} becomes then
equivalent to the MOND equation (which has the form of a modified
Poisson equation\,\footnote{We are adopting the variant of the MOND
equation which is derivable from a non-relativistic
Lagrangian~\cite{BekM84}. Henceforth we no longer indicate the
neglected remainder term $\mathcal{O}(c^{-2})$.})
\begin{equation}\label{MONDeq}
\partial_i\!\left( \mu\,g^i\right) = -4\pi G \rho \,.
\end{equation}
To account with astronomical observations we know that the Newtonian
law (without dark matter) must be valid for strong enough (though non
relativistic) gravitational fields, much above the constant
acceleration scale $a_0$. Thus $\mu\sim 1$ when $g\gg a_0$. The fact
that the acceleration scale is the relevant one is not trivial but was
found early on by Milgrom~\cite{Milg1, Milg2, Milg3}. Indeed, a
remarkable fit to many observations of rotation curves of galaxies has
been achieved by assuming that $\mu \sim g/a_0$ in the regime of weak
gravitational fields, for $g\ll a_0$. The acceleration scale $a_0$ is
empirically found to be of the order of
$10^{-10}\,\mathrm{m}\,\mathrm{s}^{-2}$ (the same numerical value, of
course, for all galaxies). Interestingly the numerical value of $a_0$
is close to the Hubble scale, $a_0\approx c\,H_0$.

Interpolating between the Newtonian and MOND regimes we have $0\leq\mu<1$,
hence the gravitational susceptibility defined by Eq.~\eqref{mu} must be
negative:
\begin{equation}\label{chi}
-1 \leq \chi < 0\,.
\end{equation}
This fact was interpreted in paper~I within a model consisting of a
pair of sub-particles with positive inertial masses but opposite
gravitational masses $\pm m$ (in analogy with the electric dipole made
of two charges $\pm q$). The negative sign in~\eqref{chi} was then
seen to reflect the fact that gravity is governed by a
\textit{negative} Coulomb law in the NR limit --- like masses attract
and unlike ones repel~\cite{Bondi57}. As a result the polarization
masses tend to \textit{increase} the magnitude of the gravitational
field, by an effect which can be referred to as gravitational
``anti-screening'', and is opposite to the usual screening of electric
charges by polarization charges in electrostatics. The negative sign
of $\chi$ was emphasized in paper~I as the main argument for viewing
MOND as a mechanism of gravitational polarization.

In principle, our task would be to justify the proportionality
relation between the polarization $\Pi^i$ and the gravitational field
$g^i$, Eq.~\eqref{Pigi}. In the model of paper~I we invoked an
equilibrium in which the distance between the sub-particles
constituting the dipole remains constant. In the present paper we have
to modify the argument because the equations of the relativistic model
(in the NR limit) are different from those of paper~I. Essentially we
shall find that when the tidal gravitational field can be neglected
there is a solution for which the dipole moments are at rest with
respect to the local matter distribution. For this solution the
internal force $F^i$ must compensate exactly for the gravitational
force. When this is realized we find that the dipole moment is aligned
with the gravitational field.

Suppose that the fluid of dipole moments fills an asymptotically flat
space-time generated by the local distribution of ordinary matter. When
choosing such asymptotically flat space-time we implicitly assume that
the isolated matter system is freely falling in the cosmological
gravitational field generated by far-distant masses in the universe. The
MOND regime will take place in the region far from the isolated system
where gravity is weak, so we expect that a good approximation is to
neglect the tidal gravitational field:
\begin{equation}\label{dijU}
\partial_{ij} U \approx 0\,.
\end{equation}
The discussion which follows is based on this approximation. More
detailed calculations taking into account the tidal gravitational
field will be performed in Sect.~\ref{secIV}.

The equation~\eqref{eveq} tells us that when~\eqref{dijU} holds
$\pi_\perp^i$ is constant or varies linearly with time. Discarding any
linear variation in time, we assume it to be constant,
\begin{equation}\label{pii}
\pi_\perp^i \approx \mathrm{const}\,,
\end{equation}
where the approximation sign $\approx$ reminds us that this is true when
the tidal gravitational field is neglected. Now we have seen that the
force $F^i$ is a function of $\pi_\perp^i$ in the NR limit, so it must
also be constant, $F^i \approx \mathrm{const}$. Similarly for the
coordinate acceleration of the dipolar particle given by
Eq.~\eqref{eom}, $d^2 x^i/d t^2 \approx \mathrm{const}$. Our basic
motive is that we shall \textit{assume}, without justification if it
were not by the existence of the solution, that the coordinate
acceleration of the dipolar particles is everywhere approximately
\textit{zero}:
\begin{equation}\label{acc0}
\frac{d^2x^i}{d t^2} \approx 0\,.
\end{equation}
This is consistent with the expectation that the dipole moments stay at
rest with respect to some averaged cosmological matter distribution on
cosmological scales. In this view the fluid of dipolar particles appears
essentially to be an immobile (static) ``ether'', weakly influenced by
the ordinary matter distribution, in agreement with the model of
paper~I. Substituting~\eqref{acc0} into the law of motion~\eqref{eom} we
obtain
\begin{equation}\label{Feq}
F^i \approx m\,g^i \,,
\end{equation}
so in this solution the internal force $F^i$ is equal to the weight
$m\,g^i$, exactly like in paper~I. Notice that
Eqs.~\eqref{acc0}--\eqref{Feq} actually say that the particle is
accelerated in a quadri-dimensional sense, because its motion is
non-geodesic (recall $m a^\mu=-F^\mu$). The quadri-force $F^\mu$ acts
like a rocket to accelerate the dipolar particle and hold it at rest
in the gravitational field.

Because $F^i$ is proportional to $\pi^i_\perp$ [see Eq.~\eqref{Fmu}],
the result~\eqref{Feq} implies that $\pi^i_\perp$ and hence the
polarization vector $\Pi^i = n\,\pi^i_\perp$ are aligned with the
gravitational field. The proportionality relation~\eqref{Pigi} is
therefore justified and we have verified the MOND
equation~\eqref{MONDeq}. Following paper~I we then look for the
expression of $F^i$ that corresponds to the specific MOND regime where
$\mu\sim g/a_0$, and obtain it in the form of an expansion when
$\pi_\perp\rightarrow 0$:
\begin{equation}\label{Fpiperp}
F^i \approx k\,m\,\pi^i_\perp \left[1 + \frac{k}{a_0} \,\pi_\perp +
\mathcal{O}\left(\pi_\perp^2\right)\right]\,.
\end{equation}
Such expansion can be viewed as a short-distance expansion in the
separation $x=\pi_\perp/m$ between the dipole's constituents. We have
posed $k=4\pi\,G\,n$ which is assumed here to be constant and
uniform.\,\footnote{Thus Eq.~\eqref{Fpiperp} depends on the space
density $n$ of particles. Note however that~\eqref{Fpiperp} represents
the value of the force at equilibrium, when the equilibrium
condition~\eqref{Feq} is satisfied. This force is computed from the
internal force $F^\mu$ defined in general by Eq.~\eqref{Fmu}, which
constitutes a definition \textit{intrinsic} to the dipole,
\textit{i.e.} valid for the single dipole moment independently of
$n$. Thus, $F^\mu$ derives from the part of the action~\eqref{S}
corresponding to a single particle, without reference to $n$.}  The
potential function $V$ from which derives the latter force reads (in
the approximation where $k$ is independent of $\pi^i_\perp$)
\begin{equation}\label{Vpiperp}
V \approx \frac{k}{2}\,\pi^2_\perp \left[1 +
\frac{2}{3}\frac{k}{a_0}\,\pi_\perp +
\mathcal{O}\left(\pi_\perp^2\right)\right]\,.
\end{equation}
It is noteworthy that this potential takes the form of an harmonic oscillator,
\textit{i.e.} is quadratic at dominant order when $\pi_\perp\rightarrow 0$
(see paper~I for discussion). In the present formalism the MOND acceleration
scale $a_0$ appears to be related \textit{via}
Eqs.~\eqref{Fpiperp}--\eqref{Vpiperp} to the properties of the internal
dipolar interaction at short distances.

To summarize, the present discussion confirms the relation between
MOND and a specific form of dipolar dark matter. Admittedly this
relation is striking --- \textit{cf} the electrostatic analogy for the
MOND equation and the polarization density~\eqref{rhopol}, the correct
sign of the susceptibility coefficient~\eqref{chi}, the harmonic
form~\eqref{Vpiperp} of the potential $V$. Dark matter could consist
of a fluid of dipole moments, polarized in the gravitational field of
ordinary galaxies in the MOND regime where $g\ll a_0$ (but inactive in
the regime of Newtonian gravitational fields where $g\gg a_0$). This
fluid of dark matter would essentially be static with respect to the
averaged cosmological matter distribution at large scales.

\section{Dipolar particles in a central gravitational field}\label{secIV}
The previous Section was restricted to the case where the tidal
gravitational field is negligible, Eq.~\eqref{dijU}. Though probably a
good approximation, this restriction hides the way the complete
equations could be integrated in more realistic situations (possibly
using numerical methods). Recall from Sect.~\ref{secIIIA} that the
dynamics of dipolar particles in the NR limit is given by the equation
of motion~\eqref{eom}, the equation of evolution~\eqref{eveq}, the
continuity equation~\eqref{continuity} and the Poisson
equation~\eqref{poisson}. In the present Section we integrate those
equations in the particular case where the gravitational field is
generated by a point mass $M$ (made of ordinary matter).

We shall assume that (1) the dipole moments are aligned with the
gravitational field, and (2) the fluid of dipolar particles is
\textit{stationary} with a purely radial flow. With those hypothesis
we shall basically show that there is a one-to-one correspondence
between the MOND function $\mu(g/a_0)$ and the potential function
$V(\pi_\perp/m)$ describing the ``internal physics'' of the dipole
moment. Note that by (1) we \textit{assume} that the dipolar medium is
polarized in the gravitational field but do not \textit{prove}
it.\,\footnote{This could be viewed as an indication that the model is
more ``effective'' than fundamental (like some models of dielectrics
where the electric polarization is parallel to the electric
field~\cite{Jackson}).} However we shall find a consistent solution
which describes the polarization state of this medium --- this partly
justifies the assumption.

The gravitational field generated by $M$ is central. We denote by
$r=\vert x^i\vert$ the radial distance to the mass, by $U(r)$ the
gravitational potential, and by $g^i=\partial_i U=U'(r)n^i$ the
gravitational field, where $n^i=x^i/r$ is the direction to the
observer and the prime denotes the derivative with respect to $r$. We
have $U'(r)<0$ and $g=\vert g^i\vert=-U'(r)$. By the assumption (1)
the dipole moment is aligned with $g^i$ so we pose
$\pi^i_\perp=-\pi_\perp n^i$, where the minus sign reflects the fact
that the moment should like $g^i$ be directed towards $M$, in
conformity with the sign of Eq.~\eqref{chi}. Here $\pi_\perp$ is the
Euclidean norm of $\pi^i_\perp$; for simplicity we keep the same name
as for the four norm~\eqref{normpiperp} because both agree in the NR
limit. Similarly the internal force $F^i$, which is parallel to the
dipole moment and in the same direction, is written as
$F^i=-F\,n^i$. Thus Eq.~\eqref{Fmu} becomes
\begin{equation}\label{F}
F=m\,\frac{d V}{d\pi_\perp}\,.
\end{equation}
By the assumption (2) the velocity of the dipolar particles is radial,
so we write it as $v^i= d x^i/d t = n^i\,d r/d t$ (such hypothesis will
partly be justified by the consistency of the solution). With these
notations Eq.~\eqref{eveq} becomes
\begin{equation}\label{pipp}
\frac{d^2\pi_\perp}{d t^2} = \pi_\perp\,U''\,,
\end{equation}
where $U''$ is the second derivative with respect to $r$, while
Eq.~\eqref{eom} reads
\begin{equation}\label{rpp}
\frac{d^2 r}{d t^2} = \frac{d V}{d \pi_\perp} - g\,.
\end{equation}

Let us now adopt an Eulerian description of the dipolar fluid. For a
stationary fluid there is no dependence on time, and in the central
potential the norms of the velocity and dipole moment depend only on the
radial distance: $v(r)$ and $\pi_\perp(r)$. Their total time derivatives
are then given by the usual Eulerian derivative as $dr/dt = v$, then
$dv/dt = v\,v'$ and $d\pi_\perp/dt = v\,\pi_\perp'$.
Equation~\eqref{pipp} is thus reduced to
\begin{equation}\label{eq1}
\gamma\,\pi_\perp' + v^2\,\pi_\perp'' = \pi_\perp\,U''\,,
\end{equation}
where $\gamma=v\,v'$ denotes the acceleration, which is itself given by
Eq.~\eqref{rpp} as
\begin{equation}\label{eq2}
\gamma = \frac{d V}{d \pi_\perp} - g\,.
\end{equation}
A consequence of the latter equation is
\begin{equation}\label{eq20}
V' = \pi_\perp'\left(\frac{v^2}{2}-U\right)'\,.
\end{equation}
We now eliminate $\gamma$ between~\eqref{eq1} and~\eqref{eq2} to obtain
\begin{equation}\label{eq3}
V' + U'\,\pi_\perp' + v^2\,\pi_\perp'' = \pi_\perp\,U''\,,
\end{equation}
which can be transformed by further manipulations [using
Eq.~\eqref{eq20}] into an equation having the form of a ``conservation''
law:
\begin{equation}\label{eq30}
(V - v^2\,\pi_\perp' + \pi_\perp\,U')'=0\,.
\end{equation}
Upon integration of this conservation law we shall obtain an arbitrary
constant (independent of $r$), \textit{i.e.} $V - v^2\,\pi_\perp' +
\pi_\perp\,U'=\mathrm{const}$. However this constant can be absorbed
into the definition of the potential $V$ (which is anyway defined up
to a constant), and we arrive at the simple result
\begin{equation}\label{V}
V = \pi_\perp\,g + v^2\,\pi_\perp'\,.
\end{equation}
Moreover we have the independent equation~\eqref{eq2} which we rewrite
using~\eqref{V} as
\begin{equation}\label{int2}
\frac{d V}{d \pi_\perp} = g + \frac{d}{d r}\biggl[ \frac{V -
\pi_\perp\,g}{2\pi_\perp'} \biggr] \,.
\end{equation}

The previous relations are valid for any central potential $U(r)$, and
we want now to specify that $U(r)$ is generated by the point mass $M$.
The density of ordinary matter in the RHS of the Poisson
equation~\eqref{poisson} is therefore $\rho=M\delta(x^i)$, where
$\delta$ is the three-dimensional Dirac function. Using also the density
of polarization~\eqref{polar}--\eqref{rhopol}, we find that the Poisson
equation becomes
\begin{equation}\label{poisseq}
\partial_i( g^i - k\,\pi^i_\perp ) = -4\pi\,G M\,\delta\,,
\end{equation}
in which we employ the notation $k=4\pi\,G\,n$ already used in
Sect.~\ref{secIIIB}. In the present case of spherical symmetry the
latter equation is equivalent to
\begin{equation}\label{gk}
g - k\,\pi_\perp = \frac{G M}{r^2}\,,
\end{equation}
where the RHS is made of the Newtonian potential for the mass $M$. From
Eq.~\eqref{Pigi} we have $k\,\pi_\perp=-\chi\,g$, where the (negative)
susceptibility coefficient $\chi$ is related to the MOND function by
$\mu=1+\chi$. Thus,
\begin{equation}\label{gmu}
g\,\mu = \frac{G M}{r^2}\,,
\end{equation}
which is nothing but the MOND equation in the case of a point mass
source, as originally postulated in Refs.~\cite{Milg1, Milg2,
Milg3}.\,\footnote{In the extreme MOND regime we have $\mu\sim g/a_0$
and the relation~\eqref{gmu} yields the famous logarithmic potential
$U\sim -\sqrt{G M a_0}\,\ln r$ explaining the flat rotation curves of
galaxies and the Tully-Fisher relation $v_\mathrm{flat}\sim (G M
a_0)^{1/4}$.} There is no surprise in recovering~\eqref{gmu} because
we have already shown on general grounds that the Poisson equation
with a dipolar source term satisfying~\eqref{Pigi} is equivalent to
the MOND equation~\eqref{MONDeq}. Equation~\eqref{gmu} could also be
directly deduced from~\eqref{MONDeq}. On the other hand we still have
the relation between $\pi_\perp$ and $g$ which we prefer to write in
terms of $\mu$ as
\begin{equation}\label{kpiperp}
k\,\pi_\perp = (1 - \mu)\,g\,.
\end{equation}

We deal next with the equation of conservation of the number of
particles~\eqref{continuity}. This equation is easy to solve because the
fluid is stationary, the velocity field is purely radial, and we are in
spherical symmetry. It suffices to say that the flow of dipolar
particles crossing the surface of the sphere $S=4\pi\,r^2$ is constant,
which means that $n\,v\,S = \widetilde{C}$ where $\widetilde{C}$
represents the number of particles passing through $S$ per unit of time
--- a constant. Combining this with Eq.~\eqref{gmu} we get
\begin{equation}\label{flow}
\frac{k\,v}{g\,\mu} = C\,,
\end{equation}
where $C$ is a constant giving the flow through $S$ per unit of central
mass $M$ (\textit{i.e.} $C=\widetilde{C}/M$). Both $C$ and $v$ are
negative for an inward flow directed toward the central mass, and
positive for an outward flow. We shall find that for the same
configuration for $\pi_\perp$, $n$ and $V$ the two solutions are
possible, and can be deduced from each other by a time reversal. For
definiteness we shall choose $C>0$.

Our equations are the first integral~\eqref{V}, the equation of
motion~\eqref{int2}, the gravitational field equation~\eqref{gmu}, the
polarization relation~\eqref{kpiperp} and the conservation of particles
equivalent to~\eqref{flow}. Thus, five equations in all for the five
unknowns $g$, $\pi_\perp$, $V$, $v$ and $k$ (where $k$ is equivalent to
the number density $n$). Note that in this counting we have considered
that the MOND function $\mu$ is known, but we have included the
potential function $V$ in the list of unknowns. Indeed we shall show
that once $\mu$ is specified (for instance $\mu=1-e^{-g/a_0}$ as chosen
below) the potential function $V$ is determined. It would seem \textit{a
priori} more natural to proceed the other way around, namely to specify
first $V$ because it is part of the more fundamental action~\eqref{S},
and only then to deduce $\mu$ which would tell which kind of MOND
phenomenology this action corresponds to. Since there will basically be
a bijective correspondence between $V$ and $\mu$, it is clear that the
two approaches are equivalent, and we can say that a choice for the
action~\eqref{S} determines $\mu$ as expected (at least under the
hypothesis of the present calculation).

From Eqs.~\eqref{kpiperp}--\eqref{flow} we obtain $v$ as
\begin{equation}\label{vpi}
v = C\,\pi_\perp\,\frac{\mu}{1-\mu}\,.
\end{equation}
Replacing it into~\eqref{V} yields the differential equation
\begin{equation}\label{pi'}
\pi_\perp' = \frac{1}{C^2\,\pi_\perp}
\left[\frac{V}{\pi_\perp}-g\right]\left(\frac{1-\mu}{\mu}\right)^2\,.
\end{equation}
Furthermore we can write~\eqref{int2} into the equivalent form
\begin{equation}\label{V'}
\frac{V'}{\pi_\perp'} = \frac{V}{\pi_\perp} +
C^2\,\pi_\perp^2\,\frac{\mu\,\mu'}{(1-\mu)^3}\,.
\end{equation}
Since $\mu$ is given, $g$ is known by Eq.~\eqref{gmu}, and the
equations~\eqref{pi'}--\eqref{V'} form a system of coupled differential
equations for $\pi_\perp$ and $V$. In a first stage the ratio between
$V$ and $\pi_\perp$ is integrated with the result
\begin{equation}\label{Vpi}
\frac{V}{\pi_\perp} = g + \frac{\mu}{1-\mu}\left[ g - \int \frac{d
g}{\mu}\right]\,.
\end{equation}
(The standard notation is used for the indefinite integral, defined
modulo an integration constant.) Then the solution for $\pi_\perp$ is
found to be
\begin{equation}\label{pisol}
\pi_\perp = \biggl(\frac{2}{C^2}\int d r \frac{1-\mu}{\mu}\left[ g -
\int \frac{d g}{\mu}\right]\biggr)^{1/2}\,.
\end{equation}
The velocity field is then given by~\eqref{vpi}, the density of dipole
moments $n=k/(4\pi G)$ follows from~\eqref{flow}, and $V$ is deduced
from~\eqref{Vpi}. As we said, for any choice of function $\mu$ we can
determine $V$ --- and \textit{vice versa}. In this sense we have
related the phenomenology of MOND to some more ``basic'' physics
associated with the description of dark matter.

Let us exemplify the previous resolution by showing the case of the MOND
function that corresponds to the simple susceptibility coefficient
$\chi=-e^{-g/a_0}$, namely
\begin{equation}\label{muex}
\mu = 1 - e^{-g/a_0}\,.
\end{equation}
This function is obviously interesting (though maybe very special)
because it is exponentially close to its Newtonian limit when
$g\rightarrow\infty$. In this case we readily integrate
Eq.~\eqref{Vpi} and find
\begin{equation}\label{Vex}
\frac{V}{\pi_\perp} = g\,e^{g/a_0} - a_0\bigl(e^{g/a_0} - 1\bigr)
\ln\bigl(e^{g/a_0} - 1\bigr)\,.
\end{equation}
Here the integration constant has been chosen in such a way that the
ratio $V/\pi_\perp$ becomes equivalent to $g$ in the Newtonian regime
$g\rightarrow\infty$ (any other choice would make it growing
exponentially). The dipole moment is then given by
\begin{equation}\label{piex}
\pi_\perp = \biggl(- \frac{2a_0}{C^2} \int_0^r d r' \,\frac{\ln\bigl(1 -
e^{-g'/a_0}\bigr)}{e^{g'/a_0} - 1}\biggr)^{1/2}\,,
\end{equation}
where $g'\equiv g(r')$. We have fixed the integration constant (in the
lower bound of the integral) in order to ensure that
$\pi_\perp\rightarrow 0$ in the Newtonian limit corresponding to
$r\rightarrow 0$. Using the fact that $g(r')$ is implicitly given by
Eq.~\eqref{gmu}, we obtain
\begin{equation}\label{piexpl}
\pi_\perp = \frac{(G M a_0)^{1/4}}{C} \left( - \int_{g/a_0}^{\infty}
\!dy\,y^{-3/2}\,e^{-y}\,\frac{1+(y-1)\,e^{-y}}{(1-e^{-y})^{5/2}}\,\ln
(1-e^{-y})\right)^{1/2}\,.
\end{equation}
\begin{figure}[t]\includegraphics[width=6.5cm]{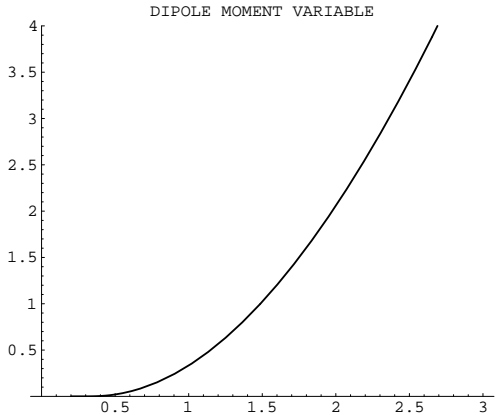}
\caption{Dipole moment $\pi_\perp$ as a function of the rescaled
  distance $\rho=r/r_0$ defined by~\eqref{r0}. On the horizontal axis
  $\rho=1$ corresponds to the transition between the Newtonian and MOND
  regimes. The unit of dipole moment on the vertical axis is $(G M
  a_0)^{1/4}/C$.}
\label{fig1}\end{figure}
\begin{figure}[t]\includegraphics[width=6.5cm]{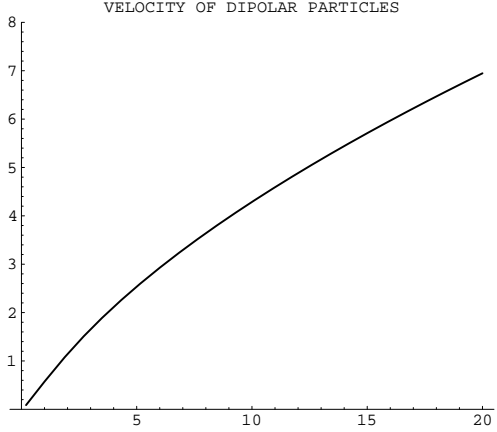}
\caption{Velocity of the dipolar particles $v$ as a function of the
  distance $\rho$. The unit of velocity on the vertical axis is $(G M
  a_0)^{1/4}$.}
\label{fig2}\end{figure}
\begin{figure}[t]\includegraphics[width=6.5cm]{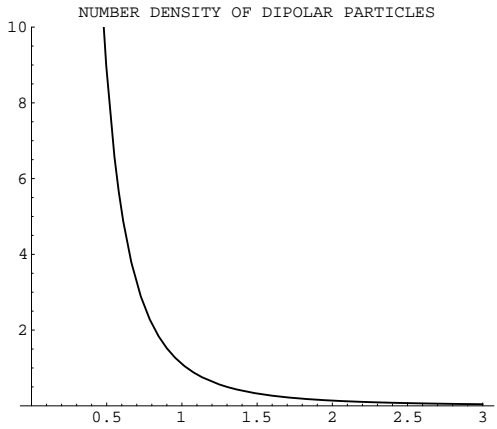}
\caption{Number density of the dipolar particles $n$ as a function of
  the distance $\rho$. The unit of number density on the vertical axis
  is $(a_0^3/G M)^{1/4}C/(4\pi G)$.}
\label{fig3}\end{figure}
\begin{figure}[t]\includegraphics[width=6.5cm]{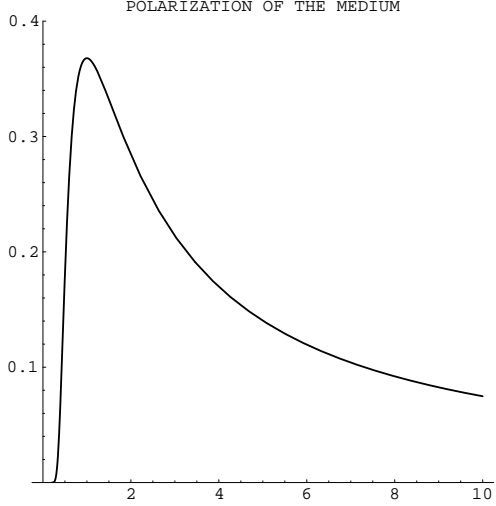}
\caption{Polarization of the medium $\Pi$ as a function of the distance
  $\rho$. The unit of polarization on the vertical axis is $a_0/(4\pi
  G)$.}
\label{fig4}\end{figure}
\begin{figure}[t]\includegraphics[width=6.5cm]{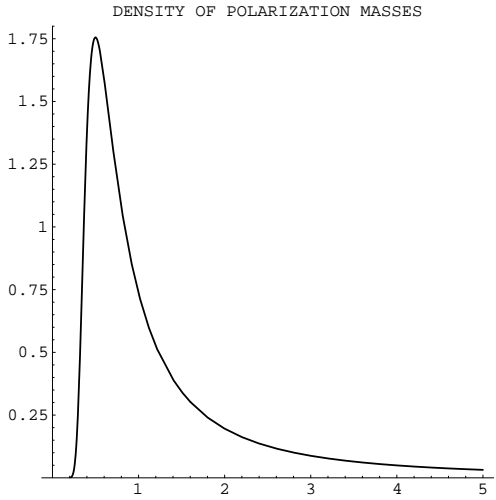}
\caption{Density of polarization masses $\rho_\mathrm{polar}$ as a
  function of the distance $\rho$. The unit of mass density on the
  vertical axis is $a_0/(4\pi G r_0)$.}
\label{fig5}\end{figure}
\begin{figure}[t]\includegraphics[width=6.5cm]{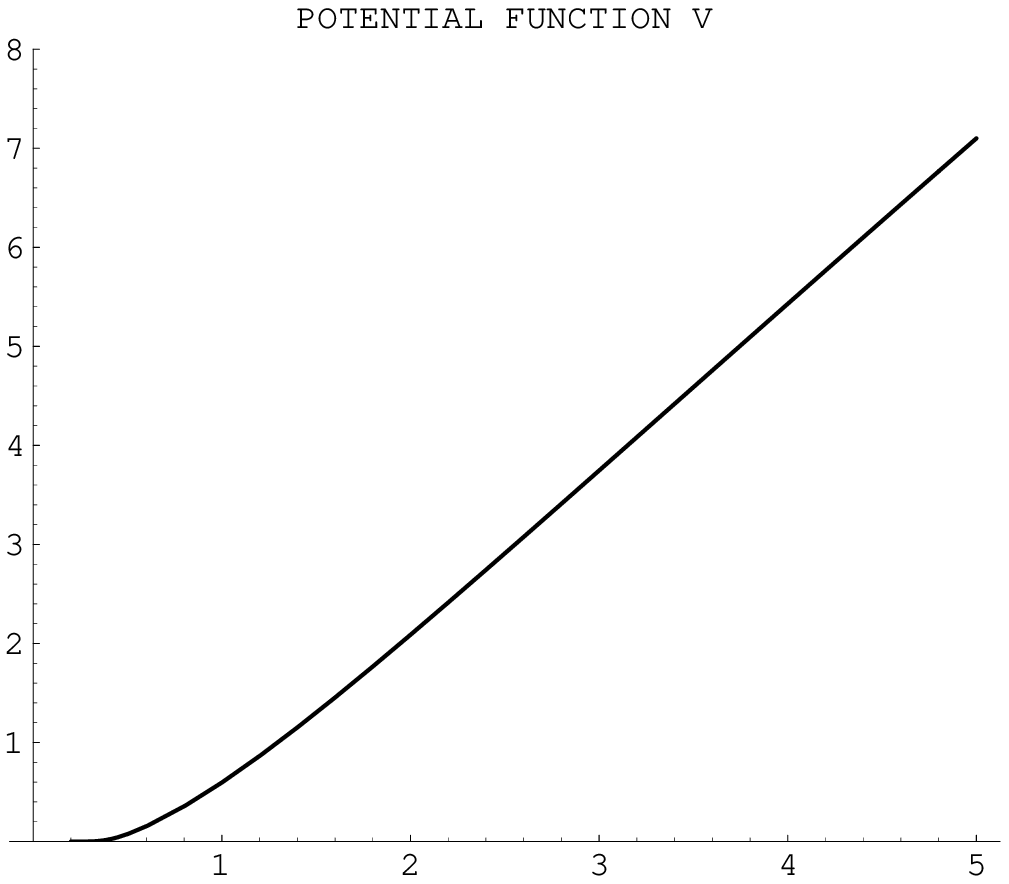}
\caption{Potential function $V$ as a function of the distance $\rho$.
The unit of $V$ on the vertical axis is $(G M a_0^5)^{1/4}/C$.}
\label{fig6}\end{figure}
In Fig.~\ref{fig1} we plot the dipole moment $\pi_\perp$ as a function
of the rescaled distance defined by $\rho\equiv r/r_0$, where $r_0$ is
the characteristic length at which the transition between the Newtonian
and MOND regimes occurs, \textit{i.e.} which is such that $g(r_0)=a_0$
and is given by
\begin{equation}\label{r0}
r_0=\left(\frac{G M}{a_0(1 - e^{-1})}\right)^{1/2}\,.
\end{equation}
In Figs.~\ref{fig2} and~\ref{fig3} we plot the velocity $v$ and the
number density $n$ of dipolar particles as functions of
$\rho=r/r_0$. We give the polarization of the medium in
Fig.~\ref{fig4}, the density of polarization masses in
Fig.~\ref{fig5}, and the potential function $V$ in Fig.~\ref{fig6}.

As we can see from Fig.~\ref{fig1} the moment $\pi_\perp$ tends
(exponentially) to zero when $r\rightarrow 0$, so the dipole moments
actually do not exist in the vicinity of the ordinary mass. However
$\pi_\perp$ diverges in the MOND domain where $r\rightarrow\infty$; more
precisely, we can check that $\pi_\perp$ behaves like $\sim r (\ln
r)^{1/2}$ when $r\rightarrow\infty$. Moreover, we see from
Fig.~\ref{fig3} that the number density $n$ of dipole moments decreases
at infinity, at the rate $\sim r^{-2}(\ln r)^{-1/2}$. To ensure the
conservation of the mass flow the velocity $v$ must then increase like
$\sim (\ln r)^{1/2}$, which is indeed what we observe in Fig.~\ref{fig2}
(recall that the particles do not obey the geodesic equation).

The fact that the dipole moment $\pi_\perp$ opens up more and more and
diverges when $r\gg r_0$ (deep in the MOND regime, far from the ordinary
matter) may appear to be surprising. However, what is important is the
\textit{density} of dipole moments or polarization $\Pi=n\,\pi_\perp$.
As we have seen the number density of dipole moments tends to zero in
the MOND domain $r\rightarrow\infty$, and as a result the polarization
(plotted by Fig.~\ref{fig4}) does decrease at large distances as
expected (at the rate $\sim r^{-1}$).\,\footnote{Figure~\ref{fig4} is
the analogue of Fig.~1 in the Newtonian-like model of paper~I (indeed
the density $n$ of dipole moments is considered to be constant in
paper~I, so the dipole moment varies like the polarization).} On the
other hand, we see that with the present choice of MOND
function~\eqref{muex} the polarization decays exponentially in the
Newtonian regime at short distances.

In the present model the polarization reflects the distribution of the
gravitational field that is induced by the fluid of dipolar particles.
The polarization shown in Fig.~\ref{fig4} is in fact given by the
analytic formula
\begin{equation}\label{Polexpr}
\Pi = \frac{g\,e^{-g/a_0}}{4\pi\,G}\,.
\end{equation}
(Notice that this formula, equivalent to~\eqref{Pigi}, could be
written directly from the standard MOND equation~\cite{Milg1, Milg2,
Milg3}; only the interpretation we propose here as the polarization of
the dipolar medium is new.)

Even more relevant than the polarization to illustrate the effect of
dark matter is the density of the polarization masses
$\rho_\mathrm{polar}$ defined by Eq.~\eqref{rhopol}. In the present case
[\textit{cf} the assumptions (1) and (2) above] we find that
\begin{equation}\label{rhopolexpl}
\rho_\mathrm{polar} = \Pi' + \frac{2\Pi}{r}\,.
\end{equation}
Figure~\ref{fig5} displays $\rho_\mathrm{polar}$ as a function of the
distance $\rho=r/r_0$. In the present model, $\rho_\mathrm{polar}$
really represents the mass density of dark matter (namely, that
density which is added to the density of ordinary matter in the source
of the Poisson equation for the Newtonian potential $U$). We can check
from Fig.~\ref{fig5} that $\rho_\mathrm{polar}$ has the correct
behaviour $\sim r^{-2}$ at infinity which reproduces the flat rotation
curves of galaxies; in more details we have
\begin{equation}\label{rhopolmond}
\rho_\mathrm{polar} \sim \frac{1}{4\pi\,r^2} \sqrt{\frac{M
a_0}{G}}~~\text{when}~~r\rightarrow\infty\,,
\end{equation}
which corresponds to the polarization mass
\begin{equation}\label{Mpolmond}
M_\mathrm{polar} \sim \sqrt{\frac{M
a_0}{G}}\,r~~\text{when}~~r\rightarrow\infty\,.
\end{equation}
Note also that a prominent feature of Fig.~\ref{fig5} is that it
predicts no accumulation of dark matter in the close vicinity of
ordinary masses, when $r\ll r_0$. Of course we meet here the natural
explanation by MOND (and therefore also by the present model) for the
absence of observed cusps of dark matter in the central regions of
galaxies. 

We have thus shown how the equations of motion and evolution of the
dipolar particles can be integrated in a specific example. Although
the solution we have considered is rather idealized --- the fluid of
dipolar particles is stationary and the dipole moments are exactly
aligned in the central gravitational field of a point
mass\,\footnote{Note that the question of the stability of this
solution against gravitational perturbations has not been
investigated.} --- it has the merit of exhibiting in details the link
between the phenomenology of MOND and the physics of dipolar
particles, specified by the potential function $V(\pi_\perp/m)$ which
enters the action~\eqref{S} [the function $V$ is plotted in
Fig.~\ref{fig6} for the particular case of MOND
function~\eqref{muex}]. 

\section{Summary and conclusions}\label{secV}

Motivated by the quasi-Newtonian model of paper~I~\cite{B06mond}, we
interpret the phenomenology of MOND as resulting from an effect of
gravitational polarization of a medium made of dipole moments aligned
in the gravitational field of ordinary masses. We propose an action
principle, based on the matter action~\eqref{S}, to describe the
dynamics of dipolar particles and the evolution of the dipole moments
in standard general relativity. The action involves a kinetic-like
term for the evolution of the dipole moment and a scalar function $V$
which is supposed to describe (at some effective level) the
non-gravitational interaction between the constituents of the dipole.

The dynamical variables are the dipolar particle's space-time position
$x^\mu$, and the dipole moment four vector $\pi^\mu$, which are varied
independently yielding the two basic equations~\eqref{Pdot}
and~\eqref{Omegadot}. Variation with respect to the metric yields the
stress-energy tensor~\eqref{Tmunu}. We find a particular class of
solutions, defined by the constraint~\eqref{Lambda1}, corresponding to
the intuitive idea of a dipole moment in equilibrium. For this class
of solutions the physical dipole moment variable is the projection
$\pi_\perp^\mu$ orthogonal to the particle's four velocity.

The non-relativistic (NR) limit of the model is investigated next. The
basic equations in this limit are (i) the equation of
motion~\eqref{eom} of the dipolar particle, (ii) an equilibrium
condition~\eqref{eveq} for the dipole moment $\pi_\perp^i$, (iii) the
conservation law~\eqref{continuity} of the number of particles $n$,
and (iv) the Poisson equation~\eqref{poisson} for the gravitational
potential $U$. The equations (i) and (ii) are different from those of
the quasi-Newtonian model of paper~I (indeed the model of paper~I
violates the equivalence principle and cannot result from the NR limit
of a general relativistic model). In addition we have the equations of
motion of ordinary massive particles~\eqref{eomord} and of
photons~\eqref{eomph}. 

Assuming that the tidal gravitational field $\partial_{ij}U$ can be
neglected, we find that there is a solution for which the dipole
moments are aligned with the gravitational field. For that solution
the Poisson equation for the gravitational field reduces to the MOND
equation like in paper~I, confirming the close relation between the
phenomenology of MOND and the dipolar dark matter.

Finally the equations in the NR limit have been integrated in the
idealized case where the fluid of dipolar particles is stationary and
the dipole moments are exactly aligned in the gravitational field
generated by a point mass $M$. The polarization of the medium is
plotted in Fig.~\ref{fig4} and the density of polarization masses
(which represents in this model the density of dark matter accumulated
around $M$) is shown in Fig.~\ref{fig5}. There is a correspondence
between the potential function $V$ in the action and the MOND function
$\mu$.

To conclude, we usually face two alternatives to the issue of dark
matter: Either accept the existence of cold dark matter particles but
which fail to reproduce in a natural way the rotation curves of
galaxies, or postulate an \textit{ad hoc} alteration of the
fundamental theory of gravity (MOND and its relativistic
extensions). In the present paper (following paper~I) we proposed a
third alternative: Keep the standard law of gravity but add to the
ordinary matter some non-standard dark matter in order to ``explain''
MOND. More precisely we invoke a mechanism of gravitational
polarization, in which the ordinary masses (galaxies) are
``anti-screened'' by polarization masses associated with gravitational
dipoles playing the role of dark matter. The dipolar dark matter
particles are in equilibrium in the gravitational field because of
their internal structure driven by some postulated non-gravitational
force (whose fundamental origin is unknown).

\acknowledgments

It is a pleasure to thank C\'edric Deffayet, Gilles Esposito-Far\`ese,
Bernard Fort, Jean-Pierre Lasota and Jean-Philippe Uzan for
interesting discussions.

\appendix

\section{Variation of the action functional}\label{appA}
In this Appendix we show how to vary the dipolar action
functional~\eqref{S}, which is of the general form
\begin{equation}\label{actionS}
S = \sum\int_{-\infty}^{+\infty} d\tau\,L\!\Bigl[g_{\mu\nu}, u^\mu,
  \pi^\mu, \dot{\pi}^\mu\Bigr] \,.
\end{equation}
The dynamical variables are the particle's space-time position
$x^\mu(\tau)$ and the dipole moment carried by the particle
$\pi^\mu(\tau)$, both depending on the proper time parametrizing the
world line and denoted by $d\tau=\sqrt{-g_{\mu\nu}d x^\mu d
x^\nu/c^2}$. The Lagrangian $L$ is a function of the dynamical
variables through the metric $g_{\mu\nu}(x)$ evaluated at
$x^\mu(\tau)$, the particle's four velocity $u^\mu = d x^\mu/d\tau$,
the dipole moment itself $\pi^\mu(\tau)$, and the covariant time
derivative of the moment $\dot{\pi}^\mu = d \pi^\mu/d \tau +
\Gamma^\mu_{\rho\sigma} u^\rho \pi^\sigma$. The four variables
$g_{\mu\nu}$, $u^\mu$, $\pi^\mu$ and $\dot{\pi}^\mu$ are considered to
be independent in Eq.~\eqref{actionS}.

We first establish a relation stating that the action~\eqref{actionS}
is a scalar \textit{vis-\`a-vis} the arbitrary infinitesimal
coordinate transformation $x'^\mu=x^\mu+\varepsilon^\mu(x)$. The
requested linear transformations of the independent variables are
\begin{equation}\left.\begin{array}{rll}
g'_{\mu\nu} &=& g_{\mu\nu} - 2
g_{\rho(\mu}\,\partial_{\nu)}\varepsilon^\rho \\[0.2cm] u'^\mu &=&
u^\mu + u^\rho \,\partial_\rho \varepsilon^\mu \\[0.2cm] \pi'^\mu &=&
\pi^\mu + \pi^\rho \,\partial_\rho \varepsilon^\mu \\[0.2cm]
\dot{\pi}'^\mu &=& \dot{\pi}^\mu + \dot{\pi}^\rho \,\partial_\rho
\varepsilon^\mu
\end{array}\right\} + \mathcal{O}(\varepsilon^2)\,,\label{transf}
\end{equation}
together with $d\tau'=d\tau$ which is a scalar. Here the prime refers
to the new values of the components of the tensor when evaluated at
the same space-time event. In the last line we have used the fact that
$\dot{\pi}^\mu$ is a four vector. The ``scalarity'' of the action
means that, under the previous coordinate transformation,
\begin{equation}\label{invS}
\int_{-\infty}^{+\infty} d\tau\,L\!\Bigl[g_{\mu\nu}, u^\mu,
  \pi^\mu, \dot{\pi}^\mu\Bigr] = \int_{-\infty}^{+\infty}
  d\tau'\,L\!\Bigl[g'_{\mu\nu}, u'^\mu, \pi'^\mu, \dot{\pi}'^\mu\Bigr]
  \,,
\end{equation}
in which the \textit{same} functional $L$ appears on both sides of the
equation. Using Eqs.~\eqref{transf} we immediately deduce the
scalarity condition
\begin{equation}\label{scalarity}
2 g_{\mu\rho}\,\frac{\partial L}{\partial g_{\nu\rho}} =
u^\nu\,\frac{\partial L}{\partial u^\mu} + \pi^\nu\,\frac{\partial
L}{\partial \pi^\mu} + \dot{\pi}^\nu\,\frac{\partial L}{\partial
\dot{\pi}^\mu}\,.
\end{equation}

We first vary the action with respect to the moment $\pi^\mu(\tau)$
subject as usual to the condition that $\pi^\mu(\pm\infty)=0$. The
variations are $\delta\pi^\mu$ and $\delta \dot{\pi}^\mu = d
\delta\pi^\mu/d \tau + \Gamma^\mu_{\rho\sigma} u^\rho
\delta\pi^\sigma$. We integrate the ordinary time derivative $d
\delta\pi^\mu/d \tau$ by part, and find that the role of the
Christoffel symbol is to make the corresponding field equation
covariant,
\begin{equation}\label{variation1}
\frac{D}{d \tau}\left(\frac{\partial L}{\partial \dot{\pi}^\mu}\right)
= \frac{\partial L}{\partial \pi^\mu}\,,
\end{equation}
where $D/d\tau$ is the covariant time derivative. The variation with
respect to the position $x^\mu(\tau)$ [satisfying
$x^\mu(\pm\infty)=0$] is more involved, due to the dependence of
$d\tau$, $u^\mu$ and $g_{\mu\nu}$ on the position. Many Christoffel
symbols and their derivatives are generated in the calculation. The
partial derivative of the Lagrangian with respect to the metric is
simplified with the help of the condition~\eqref{scalarity}. We
finally obtain a manifestly covariant equation, given by
\begin{equation}\label{variation2}
\frac{D}{d \tau}\left(\frac{\partial L}{\partial u^\mu} + u_\mu\left[
u^\rho\frac{\partial L}{\partial u^\rho} +
\dot{\pi}^\rho\frac{\partial L}{\partial \dot{\pi}^\rho} -
L\right]\right) = R^\nu_{\,\,\,\rho\mu\sigma} \pi^\rho u^\sigma
\frac{\partial L}{\partial \dot{\pi}^\nu}\,,
\end{equation}
where $R^\nu_{\,\,\,\rho\mu\sigma}$ is the Riemann curvature tensor. 

Finally we obtain the stress-energy tensor $T^{\mu\nu}$ of the fluid
of dipolar particles with number density $n$ satisfying the continuity
equation
\begin{equation}\label{continuityn}
\nabla_\mu\left(n u^\mu\right)=0\,.
\end{equation}
We thus vary the action with respect to the metric, which enters
explicitly in the first slot of $L$ in Eq.~\eqref{actionS}, and
implicitly through the proper time $d\tau$ and the covariant time
derivative $\dot{\pi}^\mu$ of the dipole moment.\,\footnote{The
variation of the Christoffel symbol present in $\dot{\pi}^\mu$ is a
tensor given by the Palatini formula as
$\delta\Gamma^\mu_{\rho\sigma}=\frac{1}{2}g^{\mu\nu}[\nabla_\rho\delta
g_{\sigma\nu}+\nabla_\sigma\delta g_{\rho\nu}-\nabla_\nu\delta
g_{\rho\sigma}]$.} In this way we obtain
\begin{eqnarray}\label{Tmunugeneral}
T^{\mu\nu} &=& n \left( 2\frac{\partial L}{\partial g_{\mu\nu}} +
u^\mu u^\nu \left[ u^\rho\frac{\partial L}{\partial u^\rho} +
\dot{\pi}^\rho\frac{\partial L}{\partial \dot{\pi}^\rho} -
L\right]\right)\nonumber\\ &+& \nabla_\rho\left\{n\left[u^{(\mu}
\pi^{\nu)}\frac{\partial L}{\partial \dot{\pi}_{\rho}} -
u^{\rho}\pi^{(\mu} \frac{\partial L}{\partial \dot{\pi}_{\nu)}} -
\pi^{\rho} u^{(\mu}\frac{\partial L}{\partial
\dot{\pi}_{\nu)}}\right]\right\}\,,
\end{eqnarray}
in which we denote $\partial L/\partial \dot{\pi}_{\rho}\equiv
g^{\rho\sigma}\partial L/\partial \dot{\pi}^{\sigma}$. The second,
dipolar-type term in this expression takes the form of the divergence
of a tensor, and has a structure similar to some related term in the
Belinfante-Rosenfeld~\cite{belinfante, rosenfeld} symmetric
stress-energy tensor of integer spin fields. An alternative expression
of~\eqref{Tmunugeneral} is readily derived with the help of the
equation of motion~\eqref{variation1} and the scalarity
condition~\eqref{scalarity}; we have [with $\partial L/\partial
u_{\rho}\equiv g^{\rho\sigma}\partial L/\partial u^{\sigma}$]
\begin{eqnarray}\label{Tmunugeneral2}
T^{\mu\nu} &=& n \left( u^{(\mu} \frac{\partial L}{\partial u_{\nu)}}
+ u^\mu u^\nu \left[ u^\rho\frac{\partial L}{\partial u^\rho} +
\dot{\pi}^\rho\frac{\partial L}{\partial \dot{\pi}^\rho} -
L\right]\right)\nonumber\\ &-& \nabla_\rho\left\{n\left[
\pi^{\rho}\frac{\partial L}{\partial \dot{\pi}_{(\mu}} -
\pi^{(\mu}\frac{\partial L}{\partial \dot{\pi}_{\rho}}\right]
u^{\nu)}\right\}\,.
\end{eqnarray}
We check that the stress-energy tensor is conserved ``on shell'',
\textit{i.e.}  when the equations of motion~\eqref{variation1}
and~\eqref{variation2} are satisfied:
\begin{equation}\label{conserved}
\nabla_\nu T^{\mu\nu} = 0\,.
\end{equation}

Finally let us make the link with the linear momenta $P_\mu$ and
$\Omega_\mu$ introduced in Eqs.~\eqref{Pmu} and~\eqref{Om}, and which
we have seen characterize entirely the dipolar particle. These momenta
appear in fact to be the conjugate momenta associated with the
dynamical variables $\pi^\mu$ and $x^\mu$ respectively, in the sense
that
\begin{subequations}\label{conjPOm}\begin{eqnarray}
P_\mu &\equiv& 2m \,\frac{\partial L}{\partial \dot{\pi}^{\mu}}\,,\\
\Omega_\mu &\equiv& \frac{\partial L}{\partial u^{\mu}} + u_\mu \left[
u^\rho\frac{\partial L}{\partial u^\rho} +
\dot{\pi}^\rho\frac{\partial L}{\partial \dot{\pi}^\rho} - L\right]\,.
\end{eqnarray}
\end{subequations}
The relation for $\Omega_\mu$ seems to be more complicated than a
simple conjugation relation, but this is due to our use of the
parametrization by the proper time $d\tau$ in the action, instead of a
parametrization which is independent of the dynamical variable
$x^\mu$. With such definitions the equations of
motion~\eqref{variation1} and~\eqref{variation2} become
\begin{subequations}\label{eom12}\begin{eqnarray}
\dot{P}_\mu &=& 2m \frac{\partial L}{\partial \pi^\mu}\,,\\
\dot{\Omega}_\mu &=& \frac{1}{2m} R_{\mu\sigma\nu\rho} u^\sigma P^\nu
\pi^\rho\,,
\end{eqnarray}
\end{subequations}
and are seen to be equivalent to the expressions derived in
Eqs.~\eqref{Pdot} and~\eqref{Omegadot} [indeed $F_\mu=-m \,\partial
L/\partial\pi^\mu$ with the explicit action~\eqref{S}]. Furthermore,
using the alternative expression~\eqref{Tmunugeneral2}, the
stress-energy tensor is obtained as
\begin{equation}\label{Tmunuapp}
T^{\mu\nu} = n \,\Omega^{(\mu} u^{\nu)} - \frac{1}{2m}
\nabla_\rho\Bigl(n \Bigl[\pi^\rho P^{(\mu} - P^\rho
\pi^{(\mu}\Bigr]u^{\nu)}\Bigr) \,.
\end{equation}
which is precisely the result given by Eq.~\eqref{Tmunu}.

\bibliography{/home/luc/Articles/ListeRef/ListeRef}

%\pagebreak
\end{document}